\documentstyle[12pt]{article}
\begin{document}
\title{\LARGE \bf
       A Multi-Configuration Mixing Approach\\
       with\\
       Symmetry-Projected Complex\\
       Hartree-Fock-Bogoliubov Determinants}
\vspace{15pt}
\author{
        E.~Bender, K.W.~Schmid and Amand~Faessler\\[10pt]
        {\it Institut f\"ur Theoretische Physik}\\
        {\it Universit\"at T\"ubingen}\\
        {\it Auf der Morgenstelle 14}\\
        {\it 72076 T\"ubingen}}
\date{}
\maketitle
\thispagestyle{empty}
\vspace{11pt}
\begin{center}
\bf Abstract
\end{center}
A multi-configuration mixing approach built on essentially complex,
symmetry-projected Har\-tree-Fock-Bogoliubov (HFB) mean fields is introduced.
The mean fields are obtained by variation after projection. The configuration
space consists out of the symmetry-projected HFB vacuum and the 
symmetry-projected two-quasiparticle excitations for even, and the 
symmetry-projected one-quasiparticle excitations for odd A systems. The
underlying complex HFB transformations are assumed to be time-reversal
invariant and axially symmetric. The model allows nuclear structure 
calculations in large model spaces with arbitrary two-body interactions. The
approach has been applied to $^{20}$Ne and $^{22}$Ne. Good agreement with the
exact shell model results and considerable improvement with respect to older
calculations, in which only real HFB transformations were admitted, is 
obtained.
\pagebreak
\section{Introduction}
\label{introduction}
\setcounter{page}{1}
\setlength{\parskip}{\baselineskip}
The shell model configuration mixing model (\cite{WHI77}, \cite{BRU77}, 
\cite{MCG80}) in general yields a very good description of at least 
the low energy phenomena in nuclear structure physics. Due to the large
dimensions of the configuration spaces, however, complete shell model 
calculations are restricted to rather small basis systems, typically
of the size of the $1s0d$ shell \cite{SCH87a}. For the description of 
many nuclear structure problems one needs much larger basis systems. 
Examples are the investigation of giant resonances, medium-heavy and 
heavy nuclei and even comparatively simple tasks like the study of negative
parity states in light even-even nuclei. In all these cases one is forced
to truncate the complete shell model expansion of the nuclear wave functions
to a manageable number of many nucleon configurations.
\par
One way to achieve this is provided by the use of variational approaches,
which leave the selection of the relevant configurations entirely to the 
dynamics of the system. The simplest models of this type are the well
known Hartree-Fock (\cite{HAR28}, \cite{FOC30}) and the more general 
Hartree-Fock-Bogoliubov (HFB) (\cite{BOG58}, \cite{BOG59a}, \cite{BOG59b}, 
\cite{RIN80}) approaches. Though the nuclear ground state is approximated 
here by one single generalized Slater determinant only, this configuration
usually accounts for a large part of the shell model expansion of the nuclear
ground state. However, in general it breaks all the symmetries required by
the many-nucleon Hamiltonian. Thus it cannot be considered as a physical
state but only as some intrinsic structure, from which the physical 
components have still to be obtained with the help of projection techniques. 
Moreover, in order to obtain really optimal solutions for each set of 
simultaneously conserved quantum numbers separately, the restoration of the
broken symmetries has to be performed before the mean field is determined 
by the variation.
\par
A whole hierarchy of such symmetry-conserving variational approaches
on the basis of HFB-type configurations have been proposed by some of us a
couple of years ago \cite{SCH84a}. They have become known as the VAMPIR 
({\bf V}ariation {\bf A}fter {\bf M}ean field {\bf P}rojection 
{\bf I}n {\bf R}ealistic model spaces) and the MONSTER ({\bf MO}del 
for handling many {\bf N}umber- and {\bf S}pin-projected 
{\bf T}wo-quasiparticle {\bf E}xcitations with {\bf R}ealistic interactions 
and model spaces) approaches.
\par
In the VAMPIR model (\cite{SCH84b}, \cite{SCH87b}) the energetically lowest
state of a particular spin-parity is approximated by a single 
symmetry-projected HFB vacuum and the underlying HFB transformation 
is determined by variation after the projection onto the desired quantum 
numbers. Excited states with the same quantum numbers can be obtained
by repeating this procedure with a new HFB test vacuum which is
constrained to be orthogonal to all the solutions already obtained.
Finally then, in this EXCITED VAMPIR model (\cite{SCH87b}, \cite{SCH86})
the residual interaction between all the obtained solutions is diagonalized.
A straightforward extension of these approaches are the Few Determinant (FED)
VAMPIR and the EXCITED FED VAMPIR models (\cite{SCH89}, \cite{PET91}), which
approximate each state not by a single, but by a linear combination of 
several non-orthogonal symmetry-projected HFB configurations, which are 
again determined by independent, successive variations. By such chains
of variational calculations the lowest few states of a given symmetry 
representation can be obtained, irrespective of their particular structure.
\par
The methods are bound to fail, however, if the complete excitation
spectrum with respect to a particular transition operator is to be 
described, like e.g., in the description of giant multipole resonances. 
If this transition operator is of one body nature, it is obviously
preferable to consider only excited states with a similar structure
like the corresponding ground (or yrast) state. One way to achieve this
is to expand the nuclear wave function around a symmetry projected
reference vacuum, which may either be the usual HFB or, e.g., a
VAMPIR solution. This is the essence of the MONSTER approach
(\cite{SCH84a}, \cite{SCH84c}, \cite{SCH87a}), in which the residual
interaction is diagonalized in the space of the symmetry-projected vacuum
and all the two-quasiparticle excitations with respect to it, if an even
system is considered while for odd systems the configuration space
is limited to the symmetry-projected one-quasiparticle configurations.
\par
Unfortunately, in all applications up to now, these models 
had to be simplified out of numerical reasons. This was achieved 
by imposing certain symmetry restrictions on the underlying
HFB transformations. Consequently, the corresponding HFB vacua
do not contain all principally possible correlations, but only a particular
part of them, which becomes more and more restricted as more symmetry 
requirements are imposed.
\par
In the first VAMPIR calculations only real, time-reversal invariant
and axially symmetric HFB transformations, which neither mix proton and
neutron states nor states of different parity, were admitted \cite{SCH84b}.
With this {\sl real} VAMPIR approach, as such calculations are called
in the following, only states in even-even nuclei with even spin and positive
parity could be described. If a MONSTER calculation is based on such
a {\sl real} VAMPIR transformation, the states with different symmetries 
(e.g., odd spins) are introduced by the configuration mixing. However, 
odd spin, or negative parity states in the same even-even nucleus,
or states in a neighbouring odd-odd nucleus may have a structure, which
differs considerably from the structure of the real reference vacuum.
Thus they cannot neccessarily be described well with the {\sl real} MONSTER
on VAMPIR approach. 
\par
A few years ago then the VAMPIR approach has been improved (\cite{SCH87b}, 
\cite{ZHE89}) by allowing essentially complex HFB transformations as well
as parity- and proton-neutron-mixing. Only time-reversal and axial symmetry
were kept. In this {\sl complex} VAMPIR approach many more nucleon 
correlations are considered and states of arbitrary spin parity in even 
mass nuclei can be described.
\par
In the present work for the first time MONSTER calculations on the basis of
such more general {\sl complex} VAMPIR transformations are performed.
The corresponding configuration spaces are considerably larger than those
of the older, more restricted real calculations and consequently a much
better description of many states is expected. Furthermore, now also
the calculations for odd-odd nuclei can be based on transformations
particularly derived for such systems, and only for the description
of odd systems one has still to rely on mean fields obtained for neighbouring
nuclei as in the older approach. However, even here the underlying
transformations are more general and consequently the configuration spaces
much larger than earlier and thus more correlations can be described.
\par
In the next section (\ref{vampirmodel}) we summarize the essential features
of the VAMPIR approach without any symmetry restrictions. We then proceed
in section (\ref{monsteronvampir}) by outlining the general formalism for 
MONSTER calculations on the basis of the corresponding VAMPIR solutions.
In section (\ref{complexmonster}) then the consequences of various symmetry 
restrictions on the underlying HFB transformation are discussed and explicit
formulas for the MONSTER approach restricted to essentially complex, but
still time-reversal invariant and axially symmetric HFB transformations are
derived. As a first test this method is applied to the two nuclei $^{20}$Ne
and $^{22}$Ne. Here only a small single particle basis, the $1s0d$-shell was
considered. This allows to compare the results not only to those of the
more restricted {\sl real} MONSTER approach but also with complete
shell model diagonalizations. This is done in section (\ref{application}). 
Finally, in section (\ref{summary}) the present work is summarized.
\section{Theory}
\label{theory}
\subsection{The VAMPIR model}
\label{vampirmodel}
The model space is defined by a finite, $D$-dimensional set of orthonormal
single particle states $ {\cal D} =\{ |i\rangle ,|k\rangle , \ldots \}_{D} $.
The indices $i,k$ are standing for the set of quantum numbers
characterizing the state. The corresponding creation and annihilation
operators are denoted by $ \{ c_i^{\dagger}, c_k^{\dagger}, \ldots \}_{D} $ and
$ \{ c_i, c_k, \ldots \}_{D} $, respectively. They obey the anti-commutation 
relations for Fermion operators. We assume that the effective many-body 
Hamiltonian appropriate for the chosen model space is known and can be 
represented by a sum of only one- and two-body terms 
\begin{equation}
  \label{phamiltonian}
       \hat{H} = \sum_{ik} t(ik) c_i^{\dagger} c_k +
         \frac{1}{4} \sum_{ikrs} v(ikrs) c_i^{\dagger} c_k^{\dagger}c_s c_r ,
\end{equation}
where $t(ik) = \langle i| \hat{t} | k \rangle $ are the one-body
matrix elements of the kinetic energy (or some single particle energies)
while $v(ikrs)= \langle ik | \hat{v} | rs - sr \rangle $ denote
the anti-symmetrized two-body matrix elements of the effective interaction.
\par
We then introduce quasiparticle cre\-a\-tion- and annihilation operators 
$a^{\dagger}$ and $a$ via the Hartree-Fock-Bogoliubov transformation 
\cite{RIN80} which in matrix notation is given by
\begin{equation}
  \label{quasiparticles}
       \left(
         \begin{array}{c} a^{\dagger}(F) \\ a(F) \end{array}
       \right)
       =
        F \left(
            \begin{array}{c} c^{\dagger} \\ c \end{array}
          \right)
       =
       \left(
         \begin{array}{cc} A^T(F) & B^T(F) \\ B^{+}(F) & A^{+}(F) \end{array}
       \right) 
       \left(
            \begin{array}{c} c^{\dagger} \\ c \end{array}
       \right).   
\end{equation}
This is the most general linear transformation conserving the Fermion
anti-com\-mu\-ta\-tion relations, provided the transformation matrix $F$ is 
chosen to be unitary. Via the inverse transformation the Hamiltonian
(\ref{phamiltonian}) can be represented in terms of the quasiparticle creation 
and annihilation operators as
\begin{equation}
  \label{qphamiltonian}
       \hat{H} = H^0(F) + \hat{H}^{11}(F) + \hat{H}^{20}(F) + \hat{H}^{22}(F)
                        + \hat{H}^{31}(F) + \hat{H}^{40}(F) ,
\end{equation}
with the upper indices denoting the number of creators and annihilators
(or vice versa), respectively. Explicit expressions for the various terms
can be found in \cite{SCH84a}. The vacuum for quasiparticle annihilators is
given by \cite{MAN75}
\begin{equation}
    \label{vacuum}
         | F \rangle =
             \left( \prod _{\alpha = 1} ^{D^{\prime}} a_{\alpha}(F) \right)
             | 0 \rangle .
\end{equation}
Here $| 0 \rangle $ is the particle vacuum and $\alpha $ enumerates the
quasiparticle states. The product runs over all quasiparticle annihilation
o\-per\-a\-tors with $ a_{\alpha} (F) | 0 \rangle \neq 0 $.
Since the HFB transformation generally mixes basis states of all
different quantum numbers, the vacuum is neither an eigenstate of the
square of the angular momentum operator, nor of its $z$-component.
Furthermore it has neither good proton nor good neutron number. 
The only symmetry which is still conserved is the so called number parity
\cite{MAN75}, i.e. the vacuum contains either only even or only odd
nucleon number components. $n$-quasiparticle states with respect to this
vacuum $ | F \rangle $ can then be defined by
\begin{equation}
  \label{n-qpstate}
          \rule[-24pt]{0pt}{14pt}
       | F \{ a^{\dagger} \}_n \rangle =
         \left\{
           \begin{array}{cl}
            | F \rangle & \mbox{for} \quad n=0 \\
            \left( \prod _{\alpha = 1} ^n a^{\dagger}_{\alpha}(F) \right)
              | F \rangle & \mbox{for} \quad n=1 ,\ldots ,D .
           \end{array}
         \right.
\end{equation}
Obviously, the $n$-qua\-si\-par\-ti\-cle states, too, do violate the
above mentioned symmetries. In order to obtain physical states, these
have to be restored before in a particular selection of such configurations
the energetically deepest solutions are determined by variation. Since any
of the $n$-quasiparticle configurations can be written as vacuum to a
particular set of quasiparticle annihilators, only HFB vacua will be
discussed in the rest of this section.
\par
The restoration of the broken symmetries is achieved with the help of
projection operators (\cite{SCH84a}, \cite{SCH87b}). Good parity $\pi$ 
is, e.g., restored by the operator
\begin{equation}
  \label{parityprojector}
       \hat{P}(\pi) = \frac{1}{2} \left[ 1 + \pi \hat{\Pi} \right] .
\end{equation}
The restoration of good proton and neutron number is equivalent to
projecting the nuclear wave function on a good mass number $A$ and
good $T_z$-component of isospin. The corresponding operators are
is given by
\begin{equation}
  \label{massprojector}
       \hat{P}(A) =
       \frac{1}{2\pi} \int\limits_{0}^{2\pi} d\phi 
         \, e^{i \phi A} \hat{S}_A(\phi )
       \quad \mbox{where} \quad \hat{S}_A(\phi )
                                \equiv e^{-i\phi\hat{A}} ,
\end{equation}
and
\begin{equation}
  \label{isospinprojector}
       \hat{P}(2T_z) =
       \frac{1}{2\pi} \int\limits_{0}^{2\pi} d\chi
         \, e^{i \chi 2T_z} \hat{S}_{2T_z}(\chi )
       \quad \mbox{where} \quad \hat{S}_{2T_z}(\chi )
                                \equiv e^{-i\chi2\hat{T_z}} ,
\end{equation}
respectively.
The desired angular momentum quantum numbers can be obtained by
an integral operator, too. It is given by
\begin{equation}
  \label{angularprojector}
       \hat{P}(IM;K) =
       \frac{2I+1}{8\pi^2} \int\limits d\Omega
         \, {D^{I^*}_{MK}} (\Omega) \hat{R}(\Omega )
\end{equation}
where $\hat{R}(\Omega)$ is the usual rotation operator and
$D^I_{MK} (\Omega)$ its representation in angular momentum eigenstates.
The integration is running over the full three Euler angles.
These projection operators commute with each other and with the Hamiltonian.
In shorthand notation we define
\begin{equation}
  \label{symmetryprojector}
       \hat{\Theta}^{AT_zI^{\pi}}_{MK} \equiv
         \hat{P}(IM;K) \hat{P}(2T_z) \hat{P}(A) \hat{P}(\pi) .
\end{equation}
\par
Physical configurations with good symmetry $S$, where $S$ represents the
quantum numbers $AT_zI^{\pi}$, are then obtained applying the projector
$\hat{\Theta}^S_{MK}$ on the above quasiparticle configurations. For
any quasiparticle vacuum $|F\rangle$, e.g., we obtain
\begin{equation}
  \label{projectedvacuum}
       |F;SM\rangle =
       \sum_{K=-I}^{I}
             \hat{\Theta}^S_{MK} |F\rangle f_K^S .
\end{equation}
Note that one has to take the sum over all angular momentum $z$-components $K$.
Otherwise the projected wave function would depend on the orientation of the
intrinsic reference frame \cite{SCH84a}.
\par
The VAMPIR approach restricts the configuration space for the lowest
(the ``yrast'') state of a given symmetry $S$ to such a single
symmetry projected vacuum. The configuration mixing degrees of
freedom $f_K^S$ as well as the underlying HFB transformation $F$ are
then determined by variation
\begin{equation}
  \label{variationeq}
       \delta E^S \equiv
         \delta \frac{\langle F;SM | \hat{H} | F;SM \rangle}
                          {\langle F;SM |  F;SM \rangle}
         = 0 .
\end{equation}
Performing this variation one obtains the optimal description
of the considered yrast state by only a single determinant.
\par
The variation leads to three sets of equations which have to be
solved self-con\-sis\-tent\-ly. The first set, resulting from the variation
with respect to the mixing degrees of freedom $f_K^S$
in (\ref{projectedvacuum}), is the diagonalization of the
Hamiltonian (\ref{phamiltonian}) in the space of the non-orthogonal
configurations $\hat{\Theta}^S_{MK}|F\rangle$
\begin{equation}
  \label{variation1}
       \sum_{K^{\prime}=-I}^{I}
        \left\{ H_{KK^{\prime}}^S - E^S N_{KK^{\prime}}^S \right\}
            f_{K^{\prime}}^S 
       = 0
\end{equation}
with the constraint
\begin{equation}
  \label{costraint1}
       (f^S)^+ N^S f^S = \bf 1 
\end{equation}
which ensures the orthonormality of the resulting states.
The Hamiltonian and the overlap matrices in these equations are given by
\begin{equation}
  \label{hnvariation1}
       \begin{array}{c}
          \rule[-12pt]{0pt}{20pt}
          H_{KK^{\prime}}^S =
              \langle F;SMK | \hat{H} |F;SMK^{\prime} \rangle
          \\
          \rule[-7pt]{0pt}{12pt}
          N_{KK^{\prime}}^S =
              \langle F;SMK | F;SMK^{\prime} \rangle
       \end{array}
       \quad \mbox{where} \quad |F;SMK \rangle \equiv
                                       \hat{\Theta}^S_{MK} |F \rangle .
\end{equation}
\par
The variation with respect to the HFB transformation leads to the second
set of equations. An elegant way to perform this variation is provided
by Thouless's Theorem \cite{THO60}. It states that any HFB vacuum 
$|F^d\rangle$ can be represented in terms of an arbitrarily chosen 
reference vacuum $|F^0\rangle$, non-orthogonal to $|F^d\rangle$, as
\begin{equation}
  \label{thouless}
       |F^d\rangle =
            \langle F^0 | F^d \rangle
            \exp \left\{ \frac{1}{2} \sum_{\mu\nu} d_{\mu\nu}
                    a_{\mu}^{\dagger}(F^0) a_{\nu}^{\dagger}(F^0) \right\}
            |F^0\rangle ,
\end{equation}
with $d$ being an anti-symmetric $D \! \times \! D$ matrix. The quasiparticles
which belong to the vacuum $|F^d\rangle$ are related to those of the
vacuum $|F^0\rangle$ via
\begin{equation}
  \label{thouless2}
        \rule{0pt}{40pt}
       \begin{array}{c}
        \left(
            \begin{array}{c}  a^{\dagger}(F^d) \\ a(F^d) \end{array}
        \right)
        =
        \left(
            \begin{array}{cc}  {L_d^{-1}}^* & -{(L_d^{-1}d)}^* \\
                               -L_d^{-1} d  &  L_d^{-1}    \end{array}
        \right)
        \left(
            \begin{array}{c}  a^{\dagger}(F^0) \\ a(F^0) \end{array}
        \right)
        \\
        \rule[-10pt]{0pt}{30pt}
        \mbox{with} \quad {\bf 1} + d^T d^* = L_d L_d^{+}.
       \end{array}
\end{equation}
Consequently the variation with respect to the matrix elements of $F$ can
be replaced by the variation with respect to the matrix elements of $d$. 
One obtains
\begin{equation}
  \label{variation2}
       \begin{array}{c}
        \rule[-18pt]{0pt}{24pt}
        \frac{\partial E^S}{\partial d_{\alpha\beta}}
        =
        \sum_{\gamma\delta}
        {(L_d^{-1})}^T_{\alpha\gamma} \;\tilde{g}_{\gamma\delta}\;
          {(L_d^{-1})}_{\delta\beta}
        = 0
        \\
        \mbox{where} \quad
        \tilde{g}_{\gamma\delta}
        \equiv
         \sum_{KK^{\prime}} {f^S_K}^*
            \langle F^d| [ \hat{H} - E^S ] \hat{\Theta}^S_{KK^{\prime}}
             a_{\gamma}^{\dagger}(F^d) a_{\delta}^{\dagger}(F^d)
             |F^d\rangle f_{K^{\prime}}^S .
        \rule[-24pt]{0pt}{24pt}
       \end{array}
\end{equation}
This equation (a sort of generalized Brillouin theorem) expresses the 
stability of the solution (\ref{projectedvacuum}) against arbitrary 
projected two-quasi\-par\-ti\-cle states with the same symmetry $S$.
\par
Since the HFB vacuum is invariant under unitary transformation of the
quasiparticle operators among themselves, these two sets of equations
are not yet sufficient to determine the HFB transformation unambiguously.
As usual we use this freedom to diagonalize the one-quasiparticle spectrum.
This yields a third set of equations
\begin{equation}
  \label{oneqpspectrum}
       \langle F| a_{\alpha}(F) \hat{H} a_{\beta}^{\dagger}(F) |F \rangle
        - \delta (\alpha ,\beta) \langle F| \hat{H} |F \rangle
        = E_{\alpha}\,\delta (\alpha ,\beta) .
\end{equation}
The eigenvalues $E_{\alpha}$ are called quasiparticle energies.
\par
These three sets form the VAMPIR variational equations
\cite{SCH87b}. Their solution yields in general already a rather
good description for the yrast states. If necessary, correlating
symmetry projected configurations can be obtained by successive
variational calculations. This is done in the FED VAMPIR approach
\cite{SCH89}. Furthermore, using orthonormality contraints, the procedure
can be easily extended to the description of excited states. This is done
in the EXCITED VAMPIR \cite{SCH87b}, and the even more general 
EXCITED FED VAMPIR \cite{SCH89} approaches. In principle, with the
help of these approaches nuclear states of arbitrary complexity can be
described. However, these methods are specifically designed, to obtain
wave functions for the lowest few states of a given symmetry $S$ only.
They can hardly be applied if, e.g., the complete excitation spectrum
with respect to some (usually one-body) transition operator is required.
For such problems it is preferable to consider only specific configurations
as they can be obtained by expanding the nuclear wave functions around
a suitable VAMPIR vacuum. This is demonstrated in the following section.
\subsection{The MONSTER on VAMPIR approach}
\label{monsteronvampir}
For a particular symmetry $S$ we define a configuration space
\begin{equation}
  \label{monsterconfigurationspace}
       \big\{ |Q;SMK\rangle \big\}
       \equiv
       \big\{ |F;SMK\rangle ,\, |F\alpha\beta;SMK\rangle ;
               \; \alpha < \beta ; \, K=-I,\ldots,+I      \big\}
\end{equation}
consisting out of the symmetry-projected vacuum 
(see also (\ref{projectedvacuum}))
\begin{equation}
  \label{monsterconfigurationspace2a}
       |F;SMK\rangle \equiv \hat{\Theta}^S_{MK} |F\rangle 
\end{equation}
and the symmetry-projected two-quasiparticle states
\begin{equation}
  \label{monsterconfigurationspace2b}
       |F\alpha\beta;SMK\rangle
       \equiv
         \hat{\Theta}^S_{MK} a^{\dagger}_{\alpha}(F) a^{\dagger}_{\beta}(F)
         |F\rangle
\end{equation}
with respect to it. Note that $S$ needs not necessarily be identical to
the symmetry for which the underlying VAMPIR mean field is obtained. In
many cases it will be sufficient to use just the VAMPIR transformation
obtained for the ground state of the considered system.
\par
The above choice of the configuration space ensures that the total 
wave functions of the excited states being linear combinations of 
these (non orthogonal) configurations
\begin{eqnarray}
  \label{2qpmonsterwavefunction}
       |\psi_i(F);SM\rangle
       &=& \left\{ \sum_K |F;SMK\rangle g^S_{0K;i}
          + \sum_{\alpha < \beta ,K}
                |F\alpha\beta ; SMK \rangle g^S_{\alpha\beta K;i} \right\}
                     \nonumber \\ 
       &=& \sum_Q |Q;SMK\rangle g^S_{QK;i} 
\end{eqnarray}
are similar in structure as the projected vacuum and can hence easily be
reached from the latter by, e.g., one-body transition operators. The
configuration mixing degrees of freedom $g^S_{QK;i}$ can then be obtained
by diagonalizing the Hamiltonian matrix 
(\ref{phamiltonian}) 
\begin{equation}
  \label{monsterhdiag}
       \sum_{Q^{\prime}}
          \left\{ H^S_{QK;Q^{\prime}K^{\prime}}
                 - E^S N^S_{QK;Q^{\prime}K^{\prime}} \right\}
                            g^S_{Q^{\prime}K^{\prime}}
       = 0
\end{equation}
being subject to the usual orthonormality constraint
\begin{equation}
  \label{monsterhdiagconstraint}
       (g^S)^{+} N^S g^S = \bf 1 .
\end{equation}
Note that the sum over $Q$ in (\ref{2qpmonsterwavefunction}) and
(\ref{monsterhdiag}) includes implicitly the sum over the $K$-components.
The Hamiltonian and overlap matrix elements entering the above equations
are given by
\begin{equation}
  \label{monster-hn-mes}
       \begin{array}{c}
        \rule{0pt}{20pt}
        H^S_{QK;Q^{\prime}K^{\prime}}
        = \langle Q;SMK| \hat{H} \hat{\Theta}^S_{KK^{\prime}}
                                     |Q^{\prime};SMK^{\prime} \rangle
        \\
        N^S_{QK;Q^{\prime}K^{\prime}}
        = \langle Q;SMK| \hat{\Theta}^S_{KK^{\prime}}
                                     |Q^{\prime};SMK^{\prime} \rangle .
        \rule{0pt}{20pt}
       \end{array}
\end{equation}
\par
As all shell model approaches, the above diagonalization 
(\ref{monsterhdiag},\ref{monsterhdiagconstraint}) yields many-nucleon 
wave functions which are contaminated by spurious center-of-mass ($C\!M$) 
excitations \cite{ELL55}, if the single particle basis $\cal D$ contains 
more than one major oscillator shell. In the MONSTER on VAMPIR approach 
these spurious admixtures are eliminated at least approximately by
a method originally proposed by Giraud \cite{GIR65}, which is based on
the following argument. For a pure harmonic oscillator basis an exact 
separation of the $C\!M$ motion is possible if all oscillator many-particle 
determinants with energies up to a certain $n\hbar\omega$ are included in the
configuration space. The $C\!M$ Hamiltonian is given in that case by
\begin{equation}
  \label{comhamiltonian}
       \hat{H}^{CM}
       = \hat{T}^{CM} + \hat{V}^{CM}
       = \frac{{\bf \hat{P}}^2}{2Am} + \frac{1}{2} Am{\omega}^2 {\bf R}^2
\end{equation}
with ${\bf R} = (1/A) \sum_{i=1}^A {\bf r}_i$ and
${\bf \hat{P}} = \sum_{i=1}^A {\bf \hat{p}}_i$, where
${\bf r}_i$ are the spatial coordinates and ${\bf \hat{p}}_i$
the momentum operators of the particles $i=1,\ldots ,A$. $\omega$ is the
harmonic oscillator constant and $m$ the nucleon mass. After diagonalizing
the $C\!M$ Hamiltonian in the $n\hbar\omega$ space, one can identify
\cite{ELL55} the spurious $C\!M$ components as $1\hbar\omega$,
$2\hbar\omega$, $\ldots$, $n\hbar\omega$ excitations and can eliminate them.
For any other basis it is not possible to choose a model space which
is complete with respect to $n\hbar\omega$ oscillator excitations. However, 
the diagonalization of the $C\!M$ Hamiltonian will produce energy eigenvalues
which are still clustered around the exact excitation energies \cite{SCH82}.
By eliminating the corresponding eigenstates at least the predominant
spurious components are removed. Therefore instead of solving equations
(\ref{monsterhdiag},\ref{monsterhdiagconstraint}), we first 
diagonalize the $C\!M$ Hamiltonian $\hat{H}^{CM}$ in the chosen 
non-orthogonal basis (\ref{monsterconfigurationspace})
\begin{equation}
  \label{monsterhcmdiag}
       \sum_{Q^{\prime}}
          \left\{ H^{CM;S}_{QK;Q^{\prime}K^{\prime}}
                 - E^{CM;S} N^S_{QK;Q^{\prime}K^{\prime}} \right\}
                            f^S_{Q^{\prime}K^{\prime}}
       = 0
\end{equation}
with the constraint
\begin{equation}
  \label{monsterhcmdiagconstraint}
       (f^S)^{+} N^S f^S = \bf 1 .
\end{equation}
The solutions with excitation energies around $1\hbar\omega$, $2\hbar\omega$,
$\ldots$ are considered as spurious. Out of those, denoted by $|f_{s}\rangle$
($s = 1, \ldots, n_s$), we can construct a projection operator
\begin{equation}
  \label{comprojector}
       \hat{P}_s = {\bf 1} - \sum_{s = 1}^{n_s} |f_s\rangle\langle f_s| ,
\end{equation}
which modifies the effective Hamiltonian (\ref{phamiltonian}) into
\begin{equation}
  \label{comprojectedh}
       \hat{\tilde{H}} = \hat{P}_s \hat{H} \hat{P}_s .
\end{equation}
The diagonalization problem (\ref{monsterhdiag},\ref{monsterhdiagconstraint})
is then solved for the modified instead of the original Hamiltonian. The
spurious states occur now at energies $E^S \sim 0$ and can hence
be easily identified.
\subsection[complex monster]
          {Restriction to complex time-reversal invariant \\
           and axially symmetric HFB transformations}
\label{complexmonster}
If no symmetry restrictions are imposed on the HFB transformations,
symmetry-projected vacua of the type (\ref{projectedvacuum}) can
be used to describe arbitrary states in arbitrary nuclei \cite{SCH87b}.
This, however, has not been achieved up to now out of numerical reasons.
Instead, for the existing numerical realisations of the VAMPIR approaches,
certain symmetry requirements were imposed on the underlying HFB 
transformations. 
\par
So, e.g., in the first VAMPIR calculations \cite{SCH84b}, axial symmetry 
and time-reversal invariance were required, parity- and proton-neutron
mixing were neglected and only real HFB transformations were admitted.
Consequently such {\sl real} VAMPIR solutions were only suitable
for even spin, positive parity states in doubly even nuclei. Performing
MONSTER calculations on top of such solutions, obviously these
restrictions are removed and states with arbitrary spin-parity in
doubly even, doubly odd and (in one-quasiparticle approximation) 
odd systems become accessible, too. However, for their calculation one 
has to rely on mean fields obtained for different spin values 
than the considered one and, for doubly odd and odd mass systems even for
neighbouring nuclei.
\par
In the more recent implementations of the VAMPIR approaches then
parity- as well as proton-neutron mixing were taken into account and
essentially complex HFB transformations were admitted. This introduces
many more correlations into the projected vacua as in the older
calculations and furthermore makes states with arbitrary spin and parity
in both doubly even and doubly odd nuclei accessible. In these
{\sl complex} VAMPIR approaches only axial symmetry and time-reversal 
invariance are kept. Thus the quasiparticle spectrum is still twofold 
degenerate and hence only states in even mass nuclei can be described
\cite{SCH87b}. The MONSTER approach, however, was up to now still
limited to the use of {\sl real} VAMPIR solutions. In the present
work for the first time {\sl complex} VAMPIR solutions have been used
in such multi-configuration mixing calculations.
\par
The mathematical apparatus of the {\sl complex} VAMPIR approach has
been described in detail elsewhere \cite{SCH87b}. In the following
we shall therefore only scetch the essential ingredients and 
concentrate on those features which are needed in a subsequent
MONSTER type calculation.
\par
Time-reversal invariance is imposed on the HFB transformation by requiring 
that with any creator $a_{\alpha}^{\dagger}$ also its time-reversed partner
$a_{\bar{\alpha}}^{\dagger} = \hat{\tau} a_{\alpha}^{\dagger} \hat{\tau}^{-1}$,
where $\hat{\tau}$ is the time-reversal operator, belongs to the same
quasiparticle representation. Axial symmetry is enforced by conserving
the $z$-component of the angular momentum. For the transformation 
coefficients of the HFB transformation (\ref{quasiparticles}) we get 
in this case the symmetries
\begin{equation}
  \label{hfbtrcoefsymm}
       \begin{array}{c}
         \rule{0pt}{20pt}
         A_{i\alpha}(F) = \delta (m_i, m_{\alpha}) A_{i \alpha}(F)
                    = \delta (m_i, m_{\alpha}) A^*_{\bar{i} \bar{\alpha}}(F),
         \\
         \rule{0pt}{20pt}
         B_{i \bar{\alpha}}(F) = \delta (m_i, m_{\alpha}) B_{i \bar{\alpha}}(F)
                    = -\delta (m_i, m_{\alpha}) B^*_{\bar{i} {\alpha}}(F),
         \\
         \rule[-10pt]{0pt}{30pt}
         A_{i \bar{\alpha}}(F) = 0 \quad \mbox{and} \quad B_{i \alpha}(F) = 0 .
       \end{array}
\end{equation}
Because of these symmetries, the vacuum (\ref{projectedvacuum}) becomes 
time-reversal invariant and furthermore an eigenstate of the $z$-component of
total angular momentum with eigenvalue zero. Consequently also all the
$n$-quasiparticle configurations (\ref{n-qpstate}) have a definite angular
momentum $z$-projection.
\par
These assumptions simplify the projection operator (\ref{symmetryprojector})
considerably. Now two of the five integrations can be performed
analytically and we obtain \cite{SCH87b}
\begin{eqnarray}
  \label{tiaxsymmprojector}
       \hat{\Theta}^S_{KK^{\prime}}
       &=& \frac{2I+1}{2} \int\limits_0^{\pi} d{\vartheta} \sin{(\vartheta)}
            d^I_{KK^{\prime}}(\vartheta) \hat{R}(\vartheta)
           \frac{1}{(2\pi)^2}
            \int\limits_{-\pi}^{\pi} d\varphi \int\limits_{0}^{\pi} d\chi
          \nonumber \\
       & & \rule{0pt}{15pt} \quad
          \left[ w(\varphi,\chi) \hat{S}(\varphi,\chi)
                  + w^*(\varphi,\chi) \hat{S}^*(\varphi,\chi)\right]
             \frac{1}{2}\left[ 1 + \pi_S \hat{\Pi} \right]
\end{eqnarray}
with $\hat{R}(\vartheta) = \exp (-i\vartheta\hat{J}_y)$ and the definitions
\begin{equation}
     \begin{array}{c}
         \rule{0pt}{24pt}
         w(\varphi,\chi) \equiv
          \exp {\left\{i\left[\frac{\varphi}{2}A
                    +\frac{\chi}{2}(2T_z)\right]\right\}} ,
     \\
         \rule{0pt}{24pt}
         \hat{S}(\varphi,\chi) \equiv
          \exp {\left\{-i\left[\frac{\varphi}{2}\hat{A}
                    +\frac{\chi}{2}(2\hat{T}_z)\right]\right\}} .
       \end{array}
\end{equation}
\par
The VAMPIR variational equations (\ref{variationeq}) consisting of
equations (\ref{variation1}, \ref{variation2} and \ref{oneqpspectrum}) 
are also simplified considerably. The diagonalization (\ref{variation1}) 
becomes now redundant since the vacuum has only one fixed angular momentum 
$z$-component $K=0$. Explicit formulas for the {\sl complex} VAMPIR 
variational equations can be found in \cite{SCH87b}. Here we only want 
to mention that the {\sl complex} VAMPIR vacuum contains still all possible
two nucleon couplings. However, it does not contain all possible four- 
and more-nucleon couplings (\cite{SCH89}, \cite{ZHE89}). This becomes
clear from the basic building blocks of the vacuum \cite{SCH87b}~:
a natural parity four nucleon state is always built either by two natural
parity or two unnatural parity pairs, but never by one natural parity 
and one unnatural parity pair. One natural parity pair and one unnatural 
parity pair always yield an unnatural four nucleon state. Analogously one 
can find the missing couplings for larger numbers of nucleons. States 
dominated by such couplings cannot be described well with the {\sl complex}
VAMPIR approach. To solve the {\sl complex} VAMPIR variational equations
one has to calculate Hamiltonian and overlap matrix elements
in between the projected vacuum. They are also needed in the
{\sl complex} MONSTER approach and therefore shown here, although
similar expressions can be found already in \cite{SCH87b}. For convenience
we introduce the abbreviations
\begin{equation}
       \begin{array}{c}
         \rule[-14pt]{0pt}{40pt}
          |F_{\lambda}\rangle \equiv
            \left\{ \begin{array}{c} |F\rangle \\ \hat{\Pi}|F\rangle
                    \end{array} \right.
          \quad \mbox{and} \quad
          a^{\dagger}(F_{\lambda}) \equiv
            \left\{ \begin{array}{c} a^{\dagger}(F) \\
                                  \hat{\Pi} \, a^{\dagger}(F) \, \hat{\Pi}^{-1}
                    \end{array} \right.
          \quad
          \left. \begin{array}{c} \mbox{for} \;\lambda =1 \\
                                 \mbox{for} \;\lambda =2
                \end{array} \right.
       \\
         \rule[-7pt]{0pt}{27pt}
         \hat{\tilde{R}}(\tilde{\Omega}) \equiv
         \hat{\tilde{R}}(\vartheta,\varphi,\chi) \equiv
                            \hat{R}(\vartheta)\hat{S}(\varphi,\chi) .
       \end{array}
\end{equation}
The projected overlap matrix element in between two vacua, which belong
to two different HFB transformations $F^1$ and $F^2$, is then given by
\begin{eqnarray}
  \label{tiaxprojoverlap}
       \langle F^1| \hat{\Theta}^S_{00} |F^2\rangle
       &=& \frac{2I+1}{2} \int\limits_0^{\pi} d{\vartheta} \sin{(\vartheta)}
            d^I_{00}(\vartheta) \frac{1}{(2\pi)^2}
            \int\limits_{-\pi}^{\pi} d\varphi \int\limits_{0}^{\pi} d\chi
          \nonumber \\
       & & \rule{0pt}{15pt}
          \Big\{ \Re e\  \Big[ w(\varphi,\chi)
                         \Big( n^{12}_{1}(\vartheta,\varphi,\chi)
                         + \pi_S n^{12}_{2}(\vartheta,\varphi,\chi) \Big) \Big]
           \Big\} ,
\end{eqnarray}
where the rotated overlap has been defined as
\begin{equation}
  \label{roto}
       n^{12}_{\lambda}(\tilde{\Omega}) \equiv
         \langle F^1|\hat{\tilde{R}}(\tilde{\Omega})
                                                     |F^2_{\lambda}\rangle .
\end{equation}
$n^{12}_{\lambda}(\tilde{\Omega})$ is calculated as described in
(\cite{SCH87b}, \cite{ONI66}, \cite{NEE83}).
General matrix elements in between two projected quasiparticle states
can then be evaluated using a generalized version of Wick's Theorem.
The only four non-vanishing elementary contractions are here
\begin{eqnarray}
  \label{elecontraction1}
       \langle F^1| a_{\alpha}(F^1) a^{\dagger}_{\beta}(F^1)
                  \hat{\tilde{R}}(\tilde{\Omega})
                                              |F^2_{\lambda} \rangle
       &\equiv & \delta (\alpha , \beta)
                 n^{12}_{\lambda}(\tilde{\Omega}) ,
       \\
  \label{elecontraction2}
       \langle F^1| a_{\beta}(F^1) a_{\alpha}(F^1)
                  \hat{\tilde{R}}(\tilde{\Omega})
                                              |F^2_{\lambda} \rangle
       &\equiv & [g^{12}_{\lambda}(\tilde{\Omega})]_{\alpha\beta}
                 n^{12}_{\lambda}(\tilde{\Omega}) ,
       \\
  \label{elecontraction3}
       \langle F^1| a_{\alpha}(F^1)
                  \hat{\tilde{R}}(\tilde{\Omega})
                  a^{\dagger}_{\beta}(F^2_{\lambda}) |F^2_{\lambda} \rangle
       &\equiv & [X^{12}_{\lambda}(\tilde{\Omega})]_{\alpha\beta}
                 n^{12}_{\lambda}(\tilde{\Omega}) ,
       \\
  \label{elecontraction4}
       \langle F^1| \hat{\tilde{R}}(\tilde{\Omega})
                  a^{\dagger}_{\alpha}(F^2_{\lambda})
                  a^{\dagger}_{\beta}(F^2_{\lambda})
                                              |F^2_{\lambda} \rangle
       &\equiv & [\tilde{g}^{12}_{\lambda}
                                    (\tilde{\Omega})]_{\alpha\beta}
                 n^{12}_{\lambda}(\tilde{\Omega}) ,
\end{eqnarray}
where
\begin{eqnarray}
  \label{X}
       X^{12}_{\lambda}(\tilde{\Omega})
       &\equiv & [A^+_{\lambda}(F^1F^2;\tilde{\Omega})]^{-1} ,
       \\
  \label{g}
       g^{12}_{\lambda}(\tilde{\Omega})
       &\equiv & B^*_{\lambda}(F^1F^2;\tilde{\Omega})
                 {X^{12}_{\lambda}}^T(\tilde{\Omega}) ,
       \\
  \label{gtilde}
       \tilde{g}^{12}_{\lambda}(\tilde{\Omega})
       &\equiv & B^T_{\lambda}(F^1F^2;\tilde{\Omega})
                 X^{12}_{\lambda}(\tilde{\Omega}) ,
\end{eqnarray}
and the rotated transformation matrices are defined as
\begin{eqnarray}
  \label{Ar}
       A_{\lambda}(F^1F^2;\tilde{\Omega})
       & \equiv &   A^+(F^1)\tilde{R}(\tilde{\Omega})A(F^2_{\lambda})
                   +B^+(F^1)\tilde{R}^*(\tilde{\Omega})B(F^2_{\lambda}) ,
       \\
  \label{Br}
       B_{\lambda}(F^1F^2;\tilde{\Omega})
       & \equiv &   B^T(F^1)\tilde{R}(\tilde{\Omega})A(F^2_{\lambda})
                   +A^T(F^1)\tilde{R}^*(\tilde{\Omega})B(F^2_{\lambda}) .
\end{eqnarray}
Here $A(F^i_{\lambda})$ and $B(F^i_{\lambda})$ are for $\lambda = 1$
just the HFB transformation matrices of the transformation $F^i$;
in the case $\lambda = 2$ these coefficients are multiplied with
the parity of the basis states. $\tilde{R}$ denotes the representation
of $\hat{\tilde{R}}$ in the chosen single particle basis $\cal D$.
Similarly the rotated energy function
\begin{equation}
  \label{roth}
       h^{n^{12}}_{\lambda}(\tilde{\Omega}) \equiv
         \langle F^1|\hat{H}\hat{\tilde{R}}(\tilde{\Omega})
                                                     |F^2_{\lambda}\rangle ,
\end{equation}
which is necessary for the calculation of the projected energy matrix element
\begin{eqnarray}
  \label{tiaxproenergy}
       \langle F^1| \hat{H} \hat{\Theta}^S_{00} |F^2\rangle
       &=& \frac{2I+1}{2} \int\limits_0^{\pi} d{\vartheta} \sin{(\vartheta)}
            d^I_{00}(\vartheta) \frac{1}{(2\pi)^2}
            \int\limits_{-\pi}^{\pi} d\varphi \int\limits_{0}^{\pi} d\chi
          \nonumber \\
       & & \rule{0pt}{15pt}
           \Big\{ \Re e\  \Big[ w(\varphi,\chi)
                    \Big( h^{n^{12}}_{1}(\vartheta,\varphi,\chi)
                     + \pi_S h^{n^{12}}_{2}(\vartheta,\varphi,\chi) \Big) \Big]
           \Big\}
\end{eqnarray}
is obtained as
\begin{equation}
  \label{hrotf}
       h^{n^{12}}_{\lambda}(\tilde{\Omega})
       =  h^{12}_{\lambda}(\tilde{\Omega})
          n^{12}_{\lambda}(\tilde{\Omega}),
\end{equation}
with
\begin{equation}
  \label{h}
       h^{12}_{\lambda}(\tilde{\Omega})
       \equiv H^0(F^1)
              + \tilde{H}^{20*}_{\lambda}(F^1F^2;\tilde{\Omega})
          + 3 \tilde{\tilde{H}} ^{40*}_{\lambda}(F^1F^2;\tilde{\Omega}) ,
\end{equation}
and
\begin{eqnarray}
       \tilde{H}^{20*}_{\lambda}(F^1F^2;\tilde{\Omega})
       \!\!\! &\equiv & \!\!\! \sum_{\alpha\beta}
                 {H_{\alpha\beta}^{20^*}}(F^1) 
                 [g^{12}_{\lambda}(\tilde{\Omega})]_{\alpha\beta} ,
       \\
       \tilde{\tilde{H}} ^{40*}_{\lambda}(F^1F^2;\tilde{\Omega})
       \!\!\! &\equiv & \!\!\! \sum_{\alpha\beta}
                 [ \tilde{H}_{\lambda}^{40*}
                            (F^1F^2;\tilde{\Omega})]_{\alpha\beta}
                 [g^{12}_{\lambda}(\tilde{\Omega})]_{\alpha\beta} ,
       \\
       {[ \tilde{H}_{\lambda}^{40*}
                             (F^1F^2;\tilde{\Omega})]}_{\alpha\beta}
       \!\!\! &\equiv & \!\!\! \sum_{\gamma\delta}
                 {H_{\alpha\beta\gamma\delta}^{40^*}}(F^1)
                 [g^{12}_{\lambda}(\tilde{\Omega})]_{\gamma\delta} ,
\end{eqnarray}
where $H^{0}$, $H^{20}$ and $H^{40}$ are Hamiltonian matrix elements of
(\ref{qphamiltonian}) in the quasiparticle representation.
\par
In case of time-reversal invariance and axial symmetry the two-quasiparticle
configurations can be written as
\begin{equation}
  \label{tiax2qp}
       \begin{array}{ll}
          \rule{0pt}{20pt}
           |F\alpha\beta\rangle \equiv
                 a^{\dagger}_{\alpha}(F) a^{\dagger}_{\beta}(F) |F\rangle ,
           & \quad
           |F\bar{\alpha}\bar{\beta}\rangle \equiv
                 \hat{\tau}|F\alpha\beta\rangle =
                 a^{\dagger}_{\bar{\alpha}}(F) a^{\dagger}_{\bar{\beta}}(F)
                                                                |F\rangle ,
          \\
          \rule{0pt}{20pt}
           |F\alpha\bar{\beta}\rangle \equiv
                 a^{\dagger}_{\alpha}(F) a^{\dagger}_{\bar{\beta}}(F)
                                                                |F\rangle ,
           & \quad
           |F{\beta}\bar{\alpha}\rangle \equiv
                 \hat{\tau}|F\alpha\bar{\beta}\rangle =
                 a^{\dagger}_{\beta}(F) a^{\dagger}_{\bar{\alpha}}(F)
                                                                |F\rangle ,
          \\
          \rule[-10pt]{0pt}{30pt}
           |F\alpha\bar{\alpha}\rangle \equiv
                 a^{\dagger}_{\alpha}(F) a^{\dagger}_{\bar{\alpha}}(F)
                                                                |F\rangle ,
           & \quad \mbox{with} \;\; \alpha < \beta \;\;
                   \mbox{and} \;\; 0 < m_{\alpha},m_{\beta} ,
       \end{array}
\end{equation}
where $m_{\alpha}$ is the angular momentum $z$-component of state $\alpha$.
Time-reversed states are considered explicitly. The configurations of the
type $|F\alpha\bar{\alpha}\rangle$ are like the vacuum invariant
under time-reversal. The total MONSTER space contains as configurations 
$|q\rangle$ the set (\ref{tiax2qp}) and the vacuum
\begin{equation}
  \label{tiaxmonsterconfigurationspace}
       \big\{ |q\rangle \big\}
       =
       \big\{|F\alpha\beta\rangle ,
             |F\bar{\alpha}\bar{\beta}\rangle ,
             |F\alpha\bar{\beta}\rangle ,
             |F\beta\bar{\alpha}\rangle ,
             |F\alpha\bar{\alpha}\rangle ,
             |F\rangle ;
             \; \alpha \!<\! \beta ,\, 0\!<\! m_{\alpha},m_{\beta}
       \big\} .
\end{equation}
Out of these configurations we construct linear combinations
which are either even or odd under time-reversal
\begin{eqnarray}
  \label{eoabbasis}
       \renewcommand{\arraystretch}{0.5}
       |F\alpha\beta ;SM\!\!\begin{array}{c}e\\o\end{array}\!\!\rangle
       \!\!\! &\equiv & \!\!\!
          \frac{1}{\sqrt{2}} \left[
           \hat{\Theta}^S_{MK_{\alpha\beta}}|F\alpha\beta\rangle
           \renewcommand{\arraystretch}{0.5}
           \! \begin{array}{c} + \\ - \end{array} \!
           \pi_S (-)^{I_S-K_{\alpha\beta}}
           \hat{\Theta}^S_{M-K_{\alpha\beta}}|F\bar{\alpha}\bar{\beta}\rangle
           \right] ,
       \\
  \label{eoabbbasis}
       \renewcommand{\arraystretch}{0.5}
       |F\alpha\bar{\beta} ;SM\!\!\begin{array}{c}e\\o\end{array}\!\!\rangle
       \!\!\! &\equiv & \!\!\!
          \frac{1}{\sqrt{2}} \left[
           \hat{\Theta}^S_{MK_{\alpha\bar{\beta}}}|F\alpha\bar{\beta}\rangle
           \renewcommand{\arraystretch}{0.5}
           \! \begin{array}{c} + \\ - \end{array} \!
           \pi_S (-)^{I_S-K_{\alpha\bar{\beta}}}
           \hat{\Theta}^S_{M-K_{\alpha\bar{\beta}}}|F\beta\bar{\alpha}\rangle
           \right] ,
\end{eqnarray}
where $K_{\alpha\beta} \equiv m_{\alpha} + m_{\beta}$ and 
$K_{\alpha\bar{\beta}} \equiv m_{\alpha} - m_{\beta}$.
The projected states of type $|F\alpha\bar{\alpha}\rangle$ are either 
even or odd depending on the spin-parity of the considered
symmetry $S$. For natural spin-parity, i.e.~$\pi_S (-)^{I_S} = +1$,
they are even, for unnatural spin-parity, i.e.~$\pi_S (-)^{I_S} = -1$,
they are odd~:
\begin{equation}
  \label{eoaabbasis}
       |F\alpha\bar{\alpha} ;SMe/o\rangle
       \equiv
          \frac{1}{2} \left[1 +\!\!/\!\!- \pi_S (-)^{I_S} \right]
           \hat{\Theta}^S_{M0}
           |F\alpha\bar{\alpha}\rangle .
\end{equation}
The same holds for the symmetry-projected vacuum $|F;SM\rangle$.
In shorthand notation we have
\begin{equation}
  \label{tiaxeomonsterconfigurationspace}
       \big\{ |{\cal Q};SMe\rangle, |{\cal Q};SMo\rangle \big\}
       \equiv
       \left\{\begin{array}{c}
               |F\alpha\beta;SMe\rangle ,
               |F\alpha\beta;SMo\rangle ,
               |F\alpha\bar{\beta};SMe\rangle ,
               \\
               |F\alpha\bar{\beta};SMo\rangle ,
               |F\alpha\bar{\alpha};SMe/o\rangle ,
               |F;SMe/o\rangle
       \end{array}\right\} .
\end{equation}
The most general wave function in this space is then given by
\begin{eqnarray}
  \label{eomonsterwavefunction}
       |\psi_i(F);SM\rangle
       &=& \sum_{\cal Q} \left\{ |{\cal Q};SMe\rangle g^{Se}_{{\cal Q};i}
                               + |{\cal Q};SMo\rangle g^{So}_{{\cal Q};i}
                               \right\}
       \\
       &=& |F;SMe/o\rangle g^{Se/o}_{0;i}
           + \sum_{\alpha \atop (0<m_{\alpha})}
             |F\alpha\bar{\alpha};SMe/o\rangle g^{Se/o}_{\alpha\bar{\alpha};i}
       \nonumber \\
       & & + \sum_{\alpha <\beta \atop (0<m_{\alpha},m_{\beta})}
             \left\{ |F\alpha\beta ;SMe\rangle g^{Se}_{\alpha\beta;i}
                   +  |F\alpha\bar{\beta};SMe\rangle
                                         g^{Se}_{\alpha\bar{\beta};i} \right.
       \nonumber \\
       & & \qquad \qquad
             \left. {} + |F\alpha\beta ;SMo\rangle g^{So}_{\alpha\beta;i}
                       + |F\alpha\bar{\beta};SMo\rangle
                                         g^{So}_{\alpha\bar{\beta};i} \right\}.
\end{eqnarray}
The expansion coefficients are obtained by diagonalizing the Hamiltonian
(or, in order to take care of the center of mass motion the modified 
Hamiltonian (\ref{comprojectedh})) according to equations 
(\ref{monsterhdiag},\ref{monsterhdiagconstraint}). For this
purpose we order the basis states $|\cal Q\rangle$ in such a way that 
first all ``even'' and then all ``odd'' configurations are listed.
In this case the hermitean Hamiltonian matrix gets the form
\begin{equation}
  \label{eohamiltonian1}
       \left( \begin{array}{cc}
              H^{ee} & H^{eo} \\ H^{oe} & H^{oo}
       \end{array}\right)
       =
       \left( \begin{array}{cc}
             \Re e\  H^{ee} & i\: \Im m\  H^{eo} \\
             -i\: \Im m\  H^{oe} & \Re e\  H^{oo}
       \end{array}\right) ,
\end{equation}
i.e., the matrix elements between two ``even'' or two ``odd'' states become
purely real, the mixed matrix elements purely imaginary. This is explicitly
shown for the two-quasiparticle states in appendix (A).
(\ref{eohamiltonian1}) can be easily brought into the form
\begin{equation}
  \label{eohamiltonian2}
       \left( \begin{array}{cc}
              {\bf 1} & 0 \\ 0 & -i\:{\bf 1}
       \end{array}\right)
       \left( \begin{array}{cc}
             \Re e\  H^{ee} & \Im m\  H^{eo} \\
              \Im m\  (H^{eo})^T & \Re e\  H^{oo}
       \end{array}\right)
       \left( \begin{array}{cc}
              {\bf 1} & 0 \\ 0 & i\:{\bf 1}
       \end{array}\right) ,
\end{equation}
which demonstrates that only a real matrix has to be diagonalized in
order to obtain the mixing coefficients $g$.
For the construction of $H^{ee}$, $H^{oo}$ and $H^{eo}$ 
it is sufficient to calculate the matrix elements
\begin{equation}
  \label{heome}
       \begin{array}{c}
          \rule[-7pt]{0pt}{14pt}
           \langle {\cal Q}^{\prime};SMe | \hat{H} | {\cal Q};SMe \rangle ,
          \\
          \rule[-7pt]{0pt}{14pt}
           \langle {\cal Q}^{\prime};SMo | \hat{H} | {\cal Q};SMo \rangle ,
          \\
          \rule[-7pt]{0pt}{14pt}
           \langle {\cal Q}^{\prime};SMe | \hat{H} | {\cal Q};SMo \rangle .
       \end{array}
\end{equation}
Since the configurations $|\cal Q\rangle$ are linear combinations
of type (\ref{eoabbasis},\ref{eoabbbasis},\ref{eoaabbasis}),
this in turn means that we have to calculate the Hamiltonian
matrix elements in between symmetry-projected states of type
(\ref{tiaxmonsterconfigurationspace}). Using the explicit
form of the symmetry projector (\ref{tiaxsymmprojector}), these are given by
\begin{eqnarray}
  \label{projmonsterme}
       \langle q^{\prime}|
            \hat{H} \hat{\Theta}^S_{K_{q^{\prime}} K_q}
            | q \rangle
       &=& \frac{2I+1}{2} \int\limits_0^{\pi} d{\vartheta} \sin{(\vartheta)}
            d^I_{K_{q^{\prime}} K_q}(\vartheta)
           \frac{1}{(2\pi)^2}
            \int\limits_{-\pi}^{\pi} d\varphi \int\limits_{0}^{\pi} d\chi
          \nonumber \\
       & &  \quad \frac{1}{2}
                \bigg\{ w(\varphi,\chi) \langle q^{\prime} |
                        \hat{H} \hat{\tilde{R}}(\tilde{\Omega})
                        | q \rangle
          \nonumber \\
       & & \quad \quad
                   {} + \pi_S \: w(\varphi,\chi) \langle q^{\prime} |
                      \hat{H} \hat{\tilde{R}}(\tilde{\Omega}) \hat{\Pi}
                        | q \rangle
          \nonumber \\
       & & \quad \quad
                   {} + w(-\varphi,-\chi) \langle q^{\prime} |
                        \hat{H} \hat{\tilde{R}}(\vartheta,-\varphi,-\chi)
                        | q \rangle
          \nonumber \\
       & & \quad \quad
                   {} + \pi_S \: w(-\varphi,-\chi) \langle q^{\prime} |
                     \hat{H} \hat{\tilde{R}}(\vartheta,-\varphi,-\chi)\hat{\Pi}
                         | q \rangle  \bigg\}.
\end{eqnarray}
For the overlap matrices one gets the analogous equations just
replacing the Hamilton operator by the unity operator $\bf 1$.
Thus finally we are left with the problem to evaluate the rotated 
Hamiltonian matrix elements appearing in equation (\ref{projmonsterme}) 
and the corresponding rotated overlaps. The matrix elements which have 
the vacuum on both sides are just the rotated energy (\ref{roth},\ref{hrotf})
and the rotated overlap (\ref{roto}). In addition there occur rotated matrix 
elements which have on one side the vacuum and on the other a 
two-quasiparticle state and rotated matrix elements in between 
two-quasiparticle states. A straightforward calculation yields for the
former
\begin{eqnarray}
  \label{roth0ab}
       \lefteqn{\langle F^1| \hat{H} \hat{\tilde{R}}(\tilde{\Omega})
                   a^{\dagger}_{\alpha}(F^2_{\lambda})
                   a^{\dagger}_{\beta}(F^2_{\lambda})
                      |F^2_{\lambda}\rangle
                = } \nonumber \\
       & & \qquad \qquad
           n^{12}_{\lambda}(\tilde{\Omega})
           \left\{ h^{12}_{\lambda}(\tilde{\Omega})
                   [\tilde{g}^{12}_{\lambda}
                                  (\tilde{\Omega})]_{\alpha\beta}
                 + [h^{02}_{\lambda}
                           (F^1F^2;\tilde{\Omega})]_{\alpha\beta}
           \right\}
\end{eqnarray}
with
\begin{eqnarray}
  \label{h02}
       [h^{02}_{\lambda}(F^1F^2;\tilde{\Omega})]_{\alpha\beta}
      \equiv 
             \Big[ {X^{12}_{\lambda}}^T(\tilde{\Omega})
                     \Big(2{H^{20}}^*(F^1) + 12 \tilde{H}_{\lambda}^{40*}
                                          (F^1F^2;\tilde{\Omega})
                      \Big)X^{12}_{\lambda}
                               (\tilde{\Omega}) \Big]_{\alpha\beta} ,
\end{eqnarray}
and
\begin{eqnarray}
  \label{rothab0}
       \lefteqn{\langle F^1| a_{\beta}(F^1) a_{\alpha}(F^1)
                         \hat{H} \hat{\tilde{R}}(\tilde{\Omega})
                         |F^2_{\lambda}\rangle
                = } \nonumber \\
       & & \qquad \qquad
           n^{12}_{\lambda}(\tilde{\Omega})
           \left\{ h^{12}_{\lambda}(\tilde{\Omega})
                   [g^{12}_{\lambda}(\tilde{\Omega})]_{\alpha\beta}
                 + [h^{20}_{\lambda}
                                (F^1F^2;\tilde{\Omega})]_{\alpha\beta}
           \right\}
\end{eqnarray}
with
\begin{eqnarray}
  \label{h20}
       \lefteqn{\rule[-10pt]{0pt}{7pt}
               [h^{20}_{\lambda}(F^1F^2;\tilde{\Omega})]_{\alpha\beta}
               \equiv } \nonumber \\
       & & \qquad
             \bigg[ 2H^{20}(F^1) + 2\tilde{H}^{22}
                                              (F^1F^2;\tilde{\Omega})
       \nonumber \\
       & &  \qquad\quad
             {}+\Big(H^{11}(F^1) - 3\tilde{H}^{31*}(F^1F^2;\tilde{\Omega})
                  \Big)g^{12}_{\lambda}(\tilde{\Omega})
       \nonumber \\
       & &   \qquad\quad
          {}- \Big( \Big(H^{11}(F^1) - 3\tilde{H}^{31*}(F^1F^2;\tilde{\Omega})
                       \Big)g^{12}_{\lambda}(\tilde{\Omega}) \Big) ^T
       \nonumber \\
       & &   \qquad\quad
             {}+g^{12}_{\lambda}(\tilde{\Omega})
                 \Big(2{H^{20}}^*(F^1)
                   + 12\tilde{H}^{40*}(F^1F^2;\tilde{\Omega})\Big)
                  g^{12}_{\lambda}(\tilde{\Omega})
             \bigg] _{\alpha\beta} ,
\end{eqnarray}
where
\begin{eqnarray}
       [\tilde{H}^{22}_{\lambda}(F^1F^2;\tilde{\Omega})]_{\alpha\beta}
       &\equiv & \sum_{\gamma\delta}
                 H_{\alpha\beta\gamma\delta}^{22}(F^1) 
                [g^{12}_{\lambda}(\tilde{\Omega})]_{\gamma\delta} ,
       \\
       {[\tilde{H}^{31*}_{\lambda}
                      (F^1F^2;\tilde{\Omega})]}_{\alpha\beta}
       &\equiv & \sum_{\gamma\delta}
                 {H_{\alpha\beta\gamma\delta}^{31^*}}(F^1) 
                 [g^{12}_{\lambda}(\tilde{\Omega})]_{\gamma\delta} ,
\end{eqnarray}
and $H^{11}$, $H^{20}$, $H^{22}$, $H^{31}$ and $H^{40}$ being the matrix
elements of the Hamiltonian (\ref{qphamiltonian}) in the quasiparticle 
representation. The corresponding rotated overlaps are just the elementary
contractions (\ref{elecontraction4}) and (\ref{elecontraction2}).
\par
The rotated Hamiltonian matrix elements in between two-quasiparticle states
on both sides are given by
\begin{eqnarray}
  \label{rothabgd}
       \lefteqn{\langle F^1| a_{\beta}(F^1) a_{\alpha}(F^1)
                         \hat{H} \hat{\tilde{R}}(\tilde{\Omega})
                         a^{\dagger}_{\gamma}(F^2) a^{\dagger}_{\delta}(F^2)
                         |F^2_{\lambda}\rangle
            =  \rule[-14pt]{0pt}{28pt} } \nonumber \\
       & & \bigg\{
           \Big( {}[g^{12}_{\lambda}(\tilde{\Omega})]_{\alpha\beta}
                 [\tilde{g}^{12}_{\lambda}(\tilde{\Omega})]_{\gamma\delta}
               + [X^{12}_{\lambda}(\tilde{\Omega})]_{\alpha\gamma}
                 [X^{12}_{\lambda}(\tilde{\Omega})]_{\beta\delta}
       \nonumber \\
       & & \qquad {}
               - [X^{12}_{\lambda}(\tilde{\Omega})]_{\alpha\delta}
                 [X^{12}_{\lambda}(\tilde{\Omega})]_{\beta\gamma}
           \Big) h^{12}_{\lambda}(\tilde{\Omega})
       \nonumber \\
       & & \quad {} \rule[-9pt]{0pt}{25pt}
               + [g^{12}_{\lambda}(\tilde{\Omega})]_{\alpha\beta}
                 [h^{02}_{\lambda}(F^1F^2;\tilde{\Omega})]_{\gamma\delta}
               + [\tilde{g}^{12}_{\lambda}(\tilde{\Omega})]_{\gamma\delta}
                 [h^{20}_{\lambda}(F^1F^2;\tilde{\Omega})]_{\alpha\beta}
       \nonumber \\
       & & \quad {}
               + [X^{12}_{\lambda}(\tilde{\Omega})]_{\alpha\gamma}
                 [h^{11}_{\lambda}(F^1F^2;\tilde{\Omega})]_{\beta\delta}
               + [X^{12}_{\lambda}(\tilde{\Omega})]_{\beta\delta}
                 [h^{11}_{\lambda}(F^1F^2;\tilde{\Omega})]_{\alpha\gamma}
       \nonumber \\
       & & \quad {}
               - [X^{12}_{\lambda}(\tilde{\Omega})]_{\alpha\delta}
                 [h^{11}_{\lambda}(F^1F^2;\tilde{\Omega})]_{\beta\gamma}
               - [X^{12}_{\lambda}(\tilde{\Omega})]_{\beta\gamma}
                 [h^{11}_{\lambda}(F^1F^2;\tilde{\Omega})]_{\alpha\delta}
       \nonumber \\
       & & \quad {}
               + [v_{\lambda}(F^1F^2;\tilde{\Omega})]_{\alpha\beta\gamma\delta}
           \bigg\} n^{12}_{\lambda}(\tilde{\Omega}) ,
\end{eqnarray}
where 
\begin{eqnarray}
  \label{h11}
       \lefteqn{ [h^{11}_{\lambda}(F^1F^2;\tilde{\Omega})]_{\alpha\beta}
       \equiv \bigg[ \Big\{ H^{11}(F^1)
                            - 3\tilde{H}^{31*}(F^1F^2;\tilde{\Omega})
               } \nonumber \\
       & & {}
             + g^{12}_{\lambda}(\tilde{\Omega})
               \left( 2{H^{20}}^*(F^1)
                    + 12 \tilde{H}_{\lambda}^{40*}
                                             (F^1F^2;\tilde{\Omega})
                   \right)
           \Big\} X^{12}_{\lambda}(\tilde{\Omega})
           \bigg] _{\alpha\beta} ,
\end{eqnarray}
and
\begin{eqnarray}
  \label{vabgd}
       \lefteqn{ [v_{\lambda}(F^1F^2;\tilde{\Omega})]
                                                 _{\alpha\beta\gamma\delta}
               \equiv \rule[-11pt]{0pt}{14pt} } \nonumber \\
       & & \qquad
           \sum_{\rho\sigma}
           \bigg\{ 4H^{22}_{\alpha\beta\rho\sigma}(F^1)
                  - \sum_{\mu\nu} \,
                       [g^{12}_{\lambda}(\tilde{\Omega})]_{\alpha\mu}
                       [g^{12}_{\lambda}(\tilde{\Omega})]_{\beta\nu}
                       24H^{40^*}_{\mu\nu\rho\sigma}(F^1)
       \nonumber \\
       & & \quad \qquad \qquad {}
         + \sum_{\nu} \left( [g^{12}_{\lambda}(\tilde{\Omega})]_{\beta\nu}
                             6H^{31^*}_{\alpha\nu\rho\sigma}(F^1)
                           - [g^{12}_{\lambda}(\tilde{\Omega})]_{\alpha\nu}
                             6H^{31^*}_{\beta\nu\rho\sigma}(F^1)
                      \right)
           \bigg\}
       \nonumber \\
       & & \qquad \qquad \quad \qquad \qquad \rule[0pt]{0pt}{15pt}
                   [X^{12}_{\lambda}(\tilde{\Omega})]_{\rho\gamma}
                   [X^{12}_{\lambda}(\tilde{\Omega})]_{\sigma\delta} .
\end{eqnarray}
The corresponding matrix elements of the rotated overlap are obtained as
\pagebreak
\begin{eqnarray}
  \label{rotnabgd}
       \lefteqn{\langle F^1| a_{\beta}(F^1) a_{\alpha}(F^1)
                         \hat{\tilde{R}}(\tilde{\Omega})
                         a^{\dagger}_{\gamma}(F^2) a^{\dagger}_{\delta}(F^2)
                         |F^2_{\lambda}\rangle
                 = \rule[-11pt]{0pt}{28pt} } \nonumber \\
       & & \quad
               \Big\{  [g^{12}_{\lambda}(\tilde{\Omega})]_{\alpha\beta}
                     [\tilde{g}^{12}_{\lambda}(\tilde{\Omega})]_{\gamma\delta}
       \nonumber \\
       & & {} \qquad
               + [X^{12}_{\lambda}(\tilde{\Omega})]_{\alpha\gamma}
                 [X^{12}_{\lambda}(\tilde{\Omega})]_{\beta\delta}
               - [X^{12}_{\lambda}(\tilde{\Omega})]_{\alpha\delta}
                 [X^{12}_{\lambda}(\tilde{\Omega})]_{\beta\gamma}
           \Big\} n^{12}_{\lambda}(\tilde{\Omega}) .
\end{eqnarray}
Various expressions occuring in the equations (\ref{roth0ab}, \ref{rothab0},
\ref{rothabgd}, and \ref{h02}, \ref{h20}) display certain symmetries related
to time-reversal. These are presented in appendix (B) and have been used in
the numerical calculation in order to reduce the actual number of matrix 
elements to be calculated \cite{BEN95}.
\par
Because of time-reversal, the vacuum as well as the two-quasiparticle
excitations with respect to it contain only components corresponding
to even mass numbers. If odd mass systems are to be considered, the
above configuration space is therefore not applicable. Instead here,
as already in the much more restricted {\sl real} MONSTER calculations,
we choose as configuration space the one-quasiparticle excitations
of an even mass vacuum
\begin{equation}
  \label{tiaxmonsterconfigurationspaceodd}
       \big\{ |q\rangle \big\}
       \equiv
       \big\{|F\alpha\rangle , |F\bar{\alpha}\rangle ,
             \, 0\!<\! m_{\alpha}
       \big\} ,
\end{equation}
where we have introduced the definitions
\begin{equation}
  \label{tiax1qp}
       |F\alpha\rangle \equiv a^{\dagger}_{\alpha}(F) |F\rangle
       \quad \mbox{and} \quad
       |F\bar{\alpha}\rangle \equiv \hat{\tau}|F\alpha\rangle =
                 a^{\dagger}_{\bar{\alpha}}(F) |F\rangle .
\end{equation}
The energy spectrum can now be calculated in complete analogy
to the even mass case. First we construct states with a good symmetry $S$
which are either even or odd under time-reversal. They have the form
\begin{equation}
  \label{eoabasis}
       \renewcommand{\arraystretch}{0.5}
       |F\alpha ;SM\!\!\begin{array}{c}e\\o\end{array}\!\!\rangle
       \equiv
          \frac{1}{\sqrt{2}} \left[
           \hat{\Theta}^S_{Mm_{\alpha}}|F\alpha\rangle
           \renewcommand{\arraystretch}{0.5}
           \! \begin{array}{c} + \\ - \end{array} \!
           \pi_S (-)^{j_S-m_{\alpha}}
           \hat{\Theta}^S_{M-m_{\alpha}}|F\bar{\alpha}\rangle
           \right] ,
\end{equation}
where small letters have been used for the angular momentum $j_S$ and 
the angular momentum $z$-component $m_{\alpha}$ to indicate the
half integer nature of these quantum numbers in odd mass nuclei.
Denoting these configurations again by 
\begin{equation}
  \label{tiaxeomonsterconfigurationspaceodd}
       \big\{ |{\cal Q};SMe\rangle, |{\cal Q};SMo\rangle \big\}
       =
       \left\{
               |F\alpha;SMe\rangle ,
               |F\alpha;SMo\rangle
       \right\} ,
\end{equation}
the most general wave function for the odd mass case takes just the same
form as for the even mass systems (\ref{eomonsterwavefunction}). 
Again the diagonalization of the Hamiltonian is reduced to a real problem 
according to equations 
(\ref{eohamiltonian1},\ref{eohamiltonian2},\ref{heome}). 
\par
The Hamiltonian matrix elements in between the one-quasiparticle
configurations $|q\rangle$ of type (\ref{tiaxmonsterconfigurationspaceodd}),
are calculated as shown in (\ref{projmonsterme}). $I$ as well as $K_q$ and 
$K_{q^{\prime}}$ of formula (\ref{projmonsterme}) have to be replaced by
the corresponding half integers. The procedure for the overlap matrix 
elements is analogous. In the end one has to calculate the rotated matrix 
elements of (\ref{projmonsterme}) in betweeen the one-quasiparticle states
(\ref{tiaxmonsterconfigurationspaceodd}). The rotated Hamiltonian matrix
elements are obtained as
\begin{eqnarray}
  \label{rothab}
       \lefteqn{\langle F^1| a_{\alpha}(F^1)
                             \hat{H} \hat{\tilde{R}}(\tilde{\Omega})
                             a^{\dagger}_{\beta}(F^2_{\lambda})
                       |F^2_{\lambda}\rangle
                = } \nonumber \\
       & & \qquad
           n^{12}_{\lambda}(\tilde{\Omega})
           \left\{ h^{12}_{\lambda}(\tilde{\Omega})
                   [X^{12}_{\lambda}(\tilde{\Omega})]_{\alpha\beta}
                 + [h^{11}_{\lambda}(F^1F^2;\tilde{\Omega})]_{\alpha\beta}
           \right\} ,
\end{eqnarray}
while the rotated overlap matrix elements are simply given by the
elementary contraction (\ref{elecontraction3}). The symmetries
of these matrix elements are displayed in appendix (B).
In general this {\sl complex} MONSTER approach for the description of odd mass
nuclei cannot be expected to have the same quality as {\sl complex} MONSTER
for even mass nuclei. Especially when the mean field properties
change quite much with the mass number, it is a clear drawback that
the one-quasiparticle states are built on a {\sl complex} VAMPIR determinant
of a neighbouring even mass nucleus. It should be stressed, however,
that because of the much richer structure of the {\sl complex} VAMPIR
solutions, the resulting configuration spaces are considerably larger
than in the {\sl real} case and hence a better description than by the latter
can be expected also for odd mass nuclei.
\par
Finally, we would like to mention, that for any particular symmetry $S$
in the following always only one vacuum $\vert F^1\rangle\equiv
\vert F^2\rangle\equiv\vert F\rangle$ will be used. Nevertheless, for
the sake of generality, all formulas in the present section have been
given for the case of two different vacua on both sides of any
matrix element.
\section{Application to $\bf ^{20}$Ne and $\bf ^{22}$Ne}
\label{application}
As a first test the above described {\sl complex} MONSTER approach
was applied to the even-even nuclei $^{20}$Ne and $^{22}$Ne using only
the $1s0d$ shell as single particle basis $\cal D$.  This rather small
basis was chosen in order to enable the comparison with complete shell model
configuration mixing calculations. The single particle energies 
$t\,(d{\,5/2}) = -4.15\mbox{ MeV}$, $t\,(s{1/2}) = -3.28\mbox{ MeV}$, 
and $t\,(d\,{3/2}) = +0.93\mbox{ MeV}$ have been taken from experiment 
\cite{AJZ86} for protons as well as neutrons. As effective two-body 
interaction the mass-dependent version of the Chung and Wildenthal force 
\cite{WIL83} ($\hat{V}(A) = \hat{V}(18) \times (18/A)^{\alpha}$) has been
used, with the only difference that instead of $\alpha = 0.3$ we took
$\alpha = 1/3$. This force has been adjusted by its authors
to many experimental data in various $sd$ nuclei and is generally
accepted to be ``the standard'' force for that model space.
The comparison of the {\sl complex} MONSTER approach with the exact
shell model calculation provides a severe test of its quality.
In addition the results are compared with those of the more restricted
{\sl real} MONSTER approach.
\par
Let us first turn our attention to $^{20}$Ne. This nucleus has two active
protons and two active neutrons in the $sd$ shell. The complete shell model
diagonalization results in 640 different states~: 46~$I^{\pi}=0^+$, 
97~$I^{\pi}=1^+$, 143~$I^{\pi}=2^+$, 129~$I^{\pi}=3^+$, 109~$I^{\pi}=4^+$,
64~$I^{\pi}=5^+$, 36~$I^{\pi}=6^+$, 12~$I^{\pi}=7^+$, and 4~$I^{\pi}=8^+$ 
states. In Fig.~1 the spectrum is plotted up to excitation energies of 
16 MeV. The {\sl complex} MONSTER approach reproduces all the 640 states 
exactly. This result is even independent of the underlying HFB 
transformation, provided the latter is essentially complex and breaks all 
the symmetries except time-reversal and axiality. The configuration mixing
degrees of freedom of the {\sl complex} MONSTER approach are here sufficient
to reproduce the complete shell model spectrum. This is understandable since
the {\sl complex} MONSTER configuration space 
(\ref{tiaxeomonsterconfigurationspace}) contains in the $sd$ shell 
altogether 277 configurations, no matter which particular (even mass)
nucleus is considered. More precisely, one has 66 configurations of type 
$|F\alpha\beta ;SMe\rangle$ and $|F\alpha\beta ;SMo\rangle$ each, 66 of type 
$|F\alpha\bar{\beta};SMe\rangle$ and $|F\alpha\bar{\beta} ;SMo \rangle$ each, 
12 of type $|F\alpha\bar{\alpha} ;SMe/o \rangle$, and the projected vacuum
$|F ;SMe/o \rangle$. Out of these, 57 configurations can contribute to spin 
$I = 0$ states, 151 are available for $I = 1$ states, 223 for $I = 2$ states, 
259 for $I = 3$ states, 275 for $I = 4$ states, and finally all 277 
are available for each of the spin $I = 5,6,7,\ldots ,14$ states. 
Since the shell model spectrum is exactly reproduced, we can conclude that
for $^{20}$Ne this configuration space does contain all the shell model
configurations.
\par
The {\sl complex} MONSTER approach is able to reproduce even those states,
which cannot be described with the {\sl complex} VAMPIR alone. These are
states which contain the ``missing couplings'', mentioned in section
(\ref{complexmonster}), as irreducible substructures. In the case of
$^{20}$Ne, e.g., the lowest $3^+$ and $5^+$ shell model states 
belong to this category. Unfortunately there is no simple ``geometrical'' 
explanation for the ``missing couplings''. To identify them, one has to 
write each state in terms of two-particle couplings using the coefficients 
of fractional parentage. For bigger basis systems this becomes very 
difficult or even impossible. For four nucleons in the $sd$ shell, however, 
this can be done easily~: out of all 29 $(d{5/2})^4$ configurations just 
two are inaccessible by the {\sl complex} VAMPIR~: the $[I^{\pi}=3^+]_{T=0}$ 
and the $[I^{\pi}=5^+]_{T=0}$ with maximal seniority, both being the dominant
structures in the corresponding yrast states (see also \cite{SCH89}). These 
states are exactly reproduced by the {\sl complex} MONSTER calculation. This
is an example which shows that couplings, which are missing in the 
{\sl complex} VAMPIR approach, can be accounted for by the configuration 
mixing of a subsequent {\sl complex} MONSTER calculation. However, we would 
like to stress ``missing couplings'' do only occur because of the restriction
to an axially and time-reversal symmetric HFB transformation. Using a HFB 
transformation without symmetry restrictions this problem is removed in the
VAMPIR approach, too.
\par
For comparison Fig.~1 also displays the spectrum obtained by the {\sl real} 
MONSTER approach built on the mean field resulting from a {\sl real} VAMPIR 
calculation for the $0^+$ ground state~: F($0^+$). In the real case the 
MONSTER approach has much less degrees of freedom than in the complex case,
and the MONSTER spectrum depends on the choice of the VAMPIR determinant
\cite{SCH84b}. First, a {\sl real} VAMPIR calculation in $sd$ shell 
accounts only for 20 variational degrees of freedom, corresponding
to the number of independent HFB transformation coefficients, while the 
{\sl complex} VAMPIR has 56, i.e. almost a factor of three more degrees of 
freedom. Second, in a {\sl real} MONSTER calculation there are altogether
only 73 configurations available~: the vacuum $|F; SM\rangle$, 
12 configurations of type $|F\alpha\bar{\alpha} ;SM\rangle$, and 30 of 
type $|F\alpha\beta ;SM\rangle$  and $|F\alpha\bar{\beta} ;SM\rangle$ each. 
21 of these configurations can contribute to $I=0$ states, 30 to $I=1$ states,
61 to $I=2$ states, 56 to $I=3$ states,  always 60 to the $I=5,7,9,11,13$ 
states, and always 73 to the $I=4,6,8,10,12,14$ states. These are only about
one fourth of those available in the {\sl complex} MONSTER approach.
Therefore it is to be expected that the energies in $^{20}$Ne obtained with
the {\sl real} MONSTER approach are higher than the exact ones, which are
reproduced by the {\sl complex} MONSTER calculation. It can be seen, however,
that the difference in energy is small. On average one has for the yrast states
a deviation of $\sim 374$ keV, i.e.~$\sim 0.9$ \% compared to the total
energy of the shell model ground state, which is $-41.412$ MeV. The largest
deviation, 624 keV, occurs for the $I^+ = 5^+$ state, the smallest for
$I^+ = 6^+$ and amounts to 135 keV.
\par
Finally we want to emphasize that the exact reproduction of the
shell model spectrum of $^{20}$ Ne by the {\sl complex} MONSTER
approach is a very stringent non-trivial test of the computer code.
\par
As second example $^{22}$Ne was studied. For this nucleus there are
in total 4206 shell model configurations~: 216~$I^{\pi} = 0^+$,
534~$I^{\pi} = 1^+$, 777~$I^{\pi} = 2^+$, 798~$I^{\pi} = 3^+$,
723~$I^{\pi} = 4^+$, 525~$I^{\pi} = 5^+$, 345~$I^{\pi} = 6^+$,
177~$I^{\pi} = 7^+$, 81~$I^{\pi} = 8^+$, 24~$I^{\pi} = 9^+$, and
6~$I^{\pi} = 10^+$ ones. In Fig.~2 for each spin only the lowest 
states resulting from the shell model diagonalization are plotted.
The energies obtained by the {\sl complex} and the {\sl real} MONSTER 
approaches, respectively, are plotted up to 16 MeV excitation energy. For
both MONSTER calculations the underlying HFB transformation was determined
by a corresponding VAMPIR calculation for the $0^+$ ground state. On 
average the {\sl complex} MONSTER reproduces the shell model energies 
of $^{22}$Ne rather well, but for most spin values not exactly,
as it was the case in $^{20}$Ne. Here, for the lower spin values
$I^{\pi} = 0^+, \ldots, 7^+$ not all the above listed shell model 
configurations are contained in the MONSTER configuration space and only
the states with angular momenta $I^{\pi} = 8^+,\ldots ,10^+$ are reproduced
exactly. Small deviations are even found for the $7^+$ states though
the number of quasiparticle configurations is here larger than the
correponding shell model dimension. This is due to the non orthogonality 
of the symmetry-projected determinants which in general causes some linear
dependencies inbetween the various MONSTER configurations. 
\par
Again the {\sl complex} MONSTER approach yields considerable improvements
with respect to its {\sl real} counterpart~: the yrast states are more
bound and many more of the excited shell model states can be described
than in the latter case.
\par
In Fig.~3 the energies of the yrast states in $^{22}$Ne as obtained
by various approaches are compared. The leftmost column shows the
exact energies obtained by the complete shell model diagonalization.
Then, from left to right, the energies calculated with the {\sl complex} 
MONSTER, the {\sl complex} VAMPIR, the {\sl real} MONSTER, and the 
{\sl real} VAMPIR are presented. Both MONSTER calculations have been
based on the corresponding VAMPIR transformations obtained for the
$0^+$ ground state. Note that with the {\sl real} VAMPIR approach only even
spin states can be described. The average deviation of the energies 
obtained by the {\sl complex} MONSTER approach from the shell model 
results is 103 keV. That is about 0.2 \% of the shell model ground 
state energy of $-58.820$ MeV. The yrast states resulting from the 
{\sl real} MONSTER approach display an average deviation of 539 keV from
the exact results. This amounts to about 0.9 \% of the shell model
ground state energy.
\par
Except for the $0^+$ ground state, where, because of the stability of the
{\sl complex} VAMPIR solution with respect to arbitrary symmetry-projected 
two-quasiparticle excitations with $K=0$, both methods yield identical
results, and except for the $9^+$ and $10^+$ state, where the latter approach
already reproduces the exact energies, the {\sl complex} MONSTER solutions are
more bound than the corresponding VAMPIR ones. This is essentially due to the
admixture of configurations with $K\neq 0$. Furthermore it becomes obvious
that the odd spin yrast states $1^+$, $3^+$, $5^+$ and $7^+$ are described 
rather poorly by {\sl complex} VAMPIR alone. One can conclude that these 
states are again dominated by components corresponding to the above discussed 
``missing couplings''. In fact, whenever the MONSTER result for an yrast state
is considerably more bound than the correponding VAMPIR result this conclusion
holds. This opens the possibility to identify ``missing couplings'' even in 
large model spaces where a decomposition of the configurations with the help 
of the coefficients of fractional parentage is not possible.
\par
The even spin states obtained with the {\sl real} VAMPIR approach are 
on average 360 keV less bound than the corresponding {\sl complex} VAMPIR 
solutions. With the {\sl real} MONSTER approach this average deviation is
reduced to 98 keV. Exceptions are again the ground state, where MONSTER
and VAMPIR solutions are again identical, and the $10^+$ state, which
is exactly reproduced by the {\sl real} VAMPIR, while the MONSTER
calculation yields 101 keV less energy. This indicates that configuration
mixing on top of the real mean field determined for the $0^+$ ground state
is not suitable to describe this high spin excitation.
\par
Finally we investigated the dependence of the {\sl complex} MONSTER
energies on the mean field determined by the preceeding {\sl complex}
VAMPIR calculation. For this purpose we have built always the full
{\sl complex} MONSTER spectrum on each of the yrast solutions 
obtained with the {\sl complex} VAMPIR approach. The same was done 
in the {\sl real} approximation. Obviously in this case only mean fields
obtained for even spin states could be used.
\par
As a typical example for the results of this investigation, Fig.~4 
displays the dependence of the energies of the five lowest $0^+$ states
in $^{22}$Ne on the spin-value used in the preceeding VAMPIR calculation. 
In addition, the leftmost column presents the energies of the three 
lowest shell model states. Solid lines denote the results of the
{\sl complex} MONSTER, dotted ones those of the {\sl real} MONSTER
approach. As can be seen, the {\sl real} MONSTER results do depend
only weekly on the underlying transformation. Here the differences
in the various mean fields seem to be not very specific and can hence
be compensated almost entirely by the configuration mixing. For the 
{\sl complex} MONSTER results this does only hold as long as the $0^+$, 
$2^+$  or $4^+$ VAMPIR transformations are taken. The higher even spin 
transformations, however, become now inadequate for the description
of the $0^+$ states. Thus one may conclude that the {\sl complex} VAMPIR
transformations display a much stronger dependence on the considered
angular momentum than the more restricted {\sl real} ones. 
As discussed above, due to ``missing couplings'', the odd spin yrast
states are only poorly described by the {\sl complex} VAMPIR approach.
This is also reflected in the bad description of the $0^+$ states,
if the MONSTER diagonalization is based on an odd-spin transformation.
\par
Similar dependencies on the underlying transformations as for the
$0^+$ states are obtained for other low spin values, too. Only for
very high angular momenta where the {\sl complex} MONSTER approach exhausts
the complete shell model spaces the results become independent of the 
underlying transformation.
\par
Finally, we would like to stress that in a MONSTER calculation
(except for spin $0^+$) the lowest yrast solution for a given
angular momentum is not necessarily obtained using the VAMPIR
transformation for this particular spin, but may result from 
a mean field derived for a neighbouring spin value. This 
was already found in {\sl real} MONSTER calculations \cite{SCH84b} 
and is due to the fact that mean field and configuration-mixing 
degrees of freedom are varied successively and not simultaneously.
\section{Summary}
\label{summary}
In the last decade a couple of variational methods have been developed
which have become known as the VAMPIR family. They all work with
symmetry-projected HFB quasiparticle vacua as test wave functions,
differ, however, in the degree of sophistication by which the underlying
HFB transformations as well as the configuration mixing is determined.
Originally, out of numerical reasons, in these methods only rather
restricted HFB transformations were admitted~: time-reversal invariance
and axial symmetry were required, parity- and proton-neutron mixing
were neglected and only {\sl real} transformation coefficients were
used. In the meantime most of these restrictions have been removed.
Only the requirement of time-reversal and axiality was kept and, e.g.,
essentially {\sl complex} transformations were allowed. By these 
generalisation many more correlations are accounted for and the range of 
applicability of these few determinant approaches was considerably
increased.
\par
Unfortunately, by construction these methods are restricted to the lowest 
few states of a given symmetry only. Very often, however, one is interested
in a complete set of particular excitations, e.g., those with respect
to a specific transition operator. For such problems it is preferable
to use a multi-configuration instead of a few determinant description. This
is done in the MONSTER approach, which expands the nuclear wave function
around the VAMPIR solution for the ground (or any other yrast) state and
obtains the excitation spectrum by diagonalizing the chosen Hamiltonian in
the space of the latter and all the symmetry-projected two-quasiparticle 
configurations with respect to it. Up to now such MONSTER calculations have
only been done using the severly symmetry-restricted {\sl real} HFB 
transformations mentioned above. In the present work, now the mathematical
formalism as well as the numerical realisation of the MONSTER approach
on the basis of the much more general {\sl complex} HFB transformations
has been developed.
\par
With respect to the previous {\sl real} implementation the 
{\sl complex} MONSTER approach has a couple of essential advantages~: first 
of all, the underlying HFB transformations account already for many more
correlations than those used in the earlier calculations. Second, for the
description of doubly odd systems one has not any more to rely on mean fields
obtained for neighbouring doubly even nuclei but can use HFB transformations
obtained for the particular system under consideration. Only for odd mass
systems, which because of the even number parity of the underlying
reference vacuum have still to be described by symmetry-projected 
one-quasiparticle states, one needs transformations from neighbouring
nuclei. Third, the configuration spaces become considerably larger than in
the more restricted {\sl real} case and consequently a more complete and more
detailed description of the nuclear spectra is to be expected.
Last but not least, the admixture of two-quasiparticle configurations 
enables the description even of such states which are dominated by 
structures which are not contained in the symmetry-projected, time-reversal
invariant HFB vacua used as test wave functions in the {\sl complex} VAMPIR
approach.
\par
As a first test the {\sl complex} MONSTER approach was applied
to the nuclei $^{20}$Ne and $^{22}$Ne. For these calculations
the $1s0d$ shell was chosen as basis. Thus the results could
not only be compared with those of the more restricted {\sl real} MONSTER
approach but also with those of exact shell model diagonalizations, which
are not available in larger model spaces. It was demonstrated that
the complete shell model spectrum of $^{20}$Ne is reproduced exactly by 
the {\sl complex} MONSTER approach. For the heavier nucleus $^{22}$Ne 
still an excellent approximation to the shell model spectrum is obtained.
So, e.g., the average deviation of the energies of yrast states from 
the exact results amounts to only 0.2\% of the shell model ground 
state energy. In both nuclei the {\sl complex} MONSTER approach yields 
considerable improvements with respect to the previous {\sl real} MONSTER
model. As compared to the above stated 0.2\%, the real calculations
give in both nuclei for the yrast energies an average deviation of 0.9\%.
\par
Already the {\sl real} MONSTER approach was a rather useful tool
for nuclear structure studies. It has been successfully applied to 
light as well as medium heavy nuclei. So, e.g., a rather nice
description of various nuclei in the mass 130 region was obtained
(\cite{HAM85}, \cite{HAM86}). Since the {\sl complex} MONSTER takes
many additional correlations into account, we expect that this
approach will develop into an even more powerful instrument
for nuclear structure investigations in large model spaces.
\par
Finally, we would like to stress that even in the {\sl complex} MONSTER
approach presented here, the underlying HFB transformations are still
symmetry-restricted. Only if the additional violation of time-reversal 
invariance and axial symmetry  is admitted, really optimal 
symmetry-projected vacua could be obtained by the VAMPIR variational 
procedure. In this case many additional correlations would be considered 
already on the mean field level. So, e.g., an unrestricted VAMPIR calculation
in an $1s0d$-basis has 552 independent variational degrees of freedom as 
compared to the 56 ones present in the {\sl complex} VAMPIR mean fields.
As a consequence, obviously also the corresponding MONSTER configuration
spaces are considerably increased and an even more detailed description
of nuclear states can be expected. Unrestricted calculations along
these lines are planned for the future.
\\[2.5\parskip]
We thank Prof.~Dr.~H.~M\" uther for performing the shell model calculations. 
This work was partly supported by the Graduiertenkolleg
``Struktur und Wechselwirkung von Hadronen und Kernen'' (DFG, Mu 705/3).
\begin{appendix}
\section*{Appendix}
\subsection*{A \hspace{5pt} Matrix elements in the ``e/o'' basis}
In this section it is shown that in the ``even-odd'' basis because
of its special properties under time-reversal only either purely real 
or purely imaginary matrix elements are obtained. As example the 
matrix elements of the Hamiltonian in between different two-quasiparticle
configurations are presented.
\pagebreak
\begin{eqnarray}
          \lefteqn{
          \langle F\alpha\beta ;SMe | \hat{H} | F\gamma\delta ;SMe \rangle }
          \nonumber \\ &=&
          \frac{1}{2} \bigg\{
          \langle F\alpha\beta | \hat{\Theta}^S_{K_{\alpha\beta}M}
                                 \hat{H}
                                 \hat{\Theta}^S_{MK_{\gamma\delta}}
                               | F\gamma\delta \rangle
          \nonumber \\ & & \qquad \qquad
          {} + (-1)^{K_{\alpha\beta} - K_{\gamma\delta}}
             \langle F\bar{\alpha}\bar{\beta}|
                                 \hat{\Theta}^S_{-K_{\alpha\beta}M}
                                 \hat{H}
                                 \hat{\Theta}^S_{M-K_{\gamma\delta}}
                               | F\bar{\gamma}\bar{\delta} \rangle
          \nonumber \\ & & \qquad
          {} + \pi_S (-1)^{I_S - K_{\gamma\delta}}
            \Big[ \langle F\alpha\beta|
                                         \hat{\Theta}^S_{K_{\alpha\beta}M}
                                         \hat{H}
                                         \hat{\Theta}^S_{M-K_{\gamma\delta}}
                                       | F\bar{\gamma}\bar{\delta} \rangle
          \nonumber \\ & & \qquad \qquad \qquad
          {} +       (-1)^{K_{\alpha\beta} - K_{\gamma\delta}}
                   \langle F\bar{\alpha}\bar{\beta}|
                                 \hat{\Theta}^S_{-K_{\alpha\beta}M}
                                 \hat{H}
                                 \hat{\Theta}^S_{MK_{\gamma\delta}}
                               | F\gamma\delta \rangle
           \Big]
           \bigg\}
          \nonumber \\ &=&
          \frac{1}{2} \bigg\{
          \langle F\alpha\beta | \hat{\Theta}^S_{K_{\alpha\beta}M}
                                 \hat{H}
                                 \hat{\Theta}^S_{MK_{\gamma\delta}}
                               | F\gamma\delta \rangle
          \nonumber \\ & & \qquad \qquad
          {} + {\langle F\alpha\beta | \hat{\Theta}^S_{K_{\alpha\beta}M}
                                 \hat{H}
                                 \hat{\Theta}^S_{MK_{\gamma\delta}}
                               | F\gamma\delta \rangle }^*
          \nonumber \\ & & \qquad
          {} + \pi_S (-1)^{I_S - K_{\gamma\delta}}
            \Big[ \langle F\alpha\beta|
                                         \hat{\Theta}^S_{K_{\alpha\beta}M}
                                         \hat{H}
                                         \hat{\Theta}^S_{M-K_{\gamma\delta}}
                                       | F\bar{\gamma}\bar{\delta} \rangle
          \nonumber \\ & & \qquad \qquad \qquad
          {} + {\langle F\alpha\beta|
                                         \hat{\Theta}^S_{K_{\alpha\beta}M}
                                         \hat{H}
                                         \hat{\Theta}^S_{M-K_{\gamma\delta}}
                                       | F\bar{\gamma}\bar{\delta} \rangle }^*
           \Big]
           \bigg\}
          \nonumber \\ &=&
              \Re e\ {\langle F\alpha\beta | \hat{\Theta}^S_{K_{\alpha\beta}M}
                                 \hat{H}
                                 \hat{\Theta}^S_{MK_{\gamma\delta}}
                               | F\gamma\delta \rangle }
          \nonumber \\ & & \qquad
          {} + \pi_S (-1)^{I_S - K_{\gamma\delta}}
              \ \Re e\  {\langle F\alpha\beta|
                                         \hat{\Theta}^S_{K_{\alpha\beta}M}
                                         \hat{H}
                                         \hat{\Theta}^S_{M-K_{\gamma\delta}}
                                       | F\bar{\gamma}\bar{\delta} \rangle } .
\end{eqnarray}
In the same way one obtains
\begin{eqnarray}
          \lefteqn{
          \langle F\alpha\beta ;SMo | \hat{H} | F\gamma\delta ;SMo \rangle }
          \nonumber \\ &=&
              \Re e\ {\langle F\alpha\beta | \hat{\Theta}^S_{K_{\alpha\beta}M}
                                 \hat{H}
                                 \hat{\Theta}^S_{MK_{\gamma\delta}}
                               | F\gamma\delta \rangle }
          \nonumber \\ & & \qquad
          {} - \pi_S (-1)^{I_S - K_{\gamma\delta}}
              \ \Re e\  {\langle F\alpha\beta|
                                         \hat{\Theta}^S_{K_{\alpha\beta}M}
                                         \hat{H}
                                         \hat{\Theta}^S_{M-K_{\gamma\delta}}
                                       | F\bar{\gamma}\bar{\delta} \rangle } ,
\end{eqnarray}
while for the mixed matrix elements one gets
\begin{eqnarray}
          \lefteqn{
          \langle F\alpha\beta ;SMe | \hat{H} | F\gamma\delta ;SMo \rangle }
          \nonumber \\ &=&
          \frac{1}{2} \bigg\{
          \langle F\alpha\beta | \hat{\Theta}^S_{K_{\alpha\beta}M}
                                 \hat{H}
                                 \hat{\Theta}^S_{MK_{\gamma\delta}}
                               | F\gamma\delta \rangle
          \nonumber \\ & & \qquad \qquad
          {} - (-1)^{K_{\alpha\beta} - K_{\gamma\delta}}
             \langle F\bar{\alpha}\bar{\beta}|
                                 \hat{\Theta}^S_{-K_{\alpha\beta}M}
                                 \hat{H}
                                 \hat{\Theta}^S_{M-K_{\gamma\delta}}
                               | F\bar{\gamma}\bar{\delta} \rangle
          \nonumber \\ & & \qquad
          {} - \pi_S (-1)^{I_S - K_{\gamma\delta}}
            \Big[ \langle F\alpha\beta|
                                         \hat{\Theta}^S_{K_{\alpha\beta}M}
                                         \hat{H}
                                         \hat{\Theta}^S_{M-K_{\gamma\delta}}
                                       | F\bar{\gamma}\bar{\delta} \rangle
          \nonumber \\ & & \qquad \qquad \qquad
          {} -       (-1)^{K_{\alpha\beta} - K_{\gamma\delta}}
                   \langle F\bar{\alpha}\bar{\beta}|
                                 \hat{\Theta}^S_{-K_{\alpha\beta}M}
                                 \hat{H}
                                 \hat{\Theta}^S_{MK_{\gamma\delta}}
                               | F\gamma\delta \rangle
           \Big]
           \bigg\}
          \nonumber \\ &=&
              i\ \Big\{
              \Im m\ {\langle F\alpha\beta | \hat{\Theta}^S_{K_{\alpha\beta}M}
                                 \hat{H}
                                 \hat{\Theta}^S_{MK_{\gamma\delta}}
                               | F\gamma\delta \rangle }
          \nonumber \\ & & \qquad
          {} - \pi_S (-1)^{I_S - K_{\gamma\delta}}
              \ \Im m\  {\langle F\alpha\beta|
                                         \hat{\Theta}^S_{K_{\alpha\beta}M}
                                         \hat{H}
                                         \hat{\Theta}^S_{M-K_{\gamma\delta}}
                                       | F\bar{\gamma}\bar{\delta} \rangle }
                \Big\} ,
\end{eqnarray}
and 
\begin{eqnarray}
          \lefteqn{
          \langle F\alpha\beta ;SMo | \hat{H} | F\gamma\delta ;SMe \rangle }
          \nonumber \\ &=&
              i\ \Big\{
              \Im m\ {\langle F\alpha\beta | \hat{\Theta}^S_{K_{\alpha\beta}M}
                                 \hat{H}
                                 \hat{\Theta}^S_{MK_{\gamma\delta}}
                               | F\gamma\delta \rangle }
          \nonumber \\ & & \qquad
          {} + \pi_S (-1)^{I_S - K_{\gamma\delta}}
              \ \Im m\  {\langle F\alpha\beta|
                                         \hat{\Theta}^S_{K_{\alpha\beta}M}
                                         \hat{H}
                                         \hat{\Theta}^S_{M-K_{\gamma\delta}}
                                       | F\bar{\gamma}\bar{\delta} \rangle }
                \Big\}
          \nonumber \\ &=&
            i\ \Big\{
              - \Im m\ {\langle F\gamma\delta |
                                 \hat{\Theta}^S_{K_{\gamma\delta}M}
                                 \hat{H}
                                 \hat{\Theta}^S_{MK_{\alpha\beta}}
                               | F\alpha\beta \rangle }
          \nonumber \\ & & \qquad
          {} + \pi_S (-1)^{I_S - K_{\alpha\beta}}
              \ \Im m\  {\langle F\gamma\delta|
                                         \hat{\Theta}^S_{K_{\gamma\delta}M}
                                         \hat{H}
                                         \hat{\Theta}^S_{M-K_{\alpha\beta}}
                                       | F\bar{\alpha}\bar{\beta} \rangle }
                \Big\}
          \nonumber \\ &=&
          {\langle F\gamma\delta ;SMe|\hat{H}|F\alpha\beta ;SMo \rangle}^* ,
\end{eqnarray}
respectively.
\subsection*{B \hspace{5pt} Symmetries}
In this appendix some symmetry relations for matrix elements needed 
in {\sl complex} MONSTER calculations are given. We restrict ourselves
here to the Hamiltonian. For the overlap matrix elements similar 
expressions are valid.
\par
For the rotated matrix elements of type
\begin{equation}
  \label{sym1roth0ab}
       [h_{\lambda}(F^1F^2;\vartheta,\varphi,\chi)]_{0\alpha\beta}
       \equiv
       \langle F^1|
                   \hat{H} \hat{\tilde{R}}(\vartheta,\varphi,\chi)
                   a^{\dagger}_{\alpha}(F^2) a^{\dagger}_{\beta}(F^2)
              |F^2_{\lambda}\rangle
\end{equation}
one gets
\begin{eqnarray}
  \label{sym2roth0ab}
    {[h_{\lambda}(F^1F^2;\vartheta,-\varphi,-\chi)]}_{0\alpha\beta}
       & = &
    {[h_{\lambda}(F^1F^2;\vartheta,\varphi,\chi)]}^*_{0\bar{\alpha}\bar{\beta}}
      \nonumber \\
    {[h_{\lambda}(F^1F^2;\vartheta,-\varphi,-\chi)]}_{0\alpha\bar{\beta}}
       & = &
    -{[h_{\lambda}(F^1F^2;\vartheta,\varphi,\chi)]}^*_{0\bar{\alpha}\beta} ,
\end{eqnarray}
while for the hermitean conjugate matrix elements
\begin{equation}
  \label{sym1rothab0}
       [h_{\lambda}(F^1F^2;\vartheta,\varphi,\chi)]_{\alpha\beta 0}
       \equiv
       \langle F^1|
                   a_{\beta}(F^1) a_{\alpha}(F^1)
                   \hat{H} \hat{\tilde{R}}(\vartheta,\varphi,\chi)
              |F^2_{\lambda}\rangle
\end{equation}
one obtains
\begin{eqnarray}
  \label{sym2rothab0}
    {[h_{\lambda}(F^1F^2;\vartheta,-\varphi,-\chi)]}_{\alpha\beta 0}
       & = &
    {[h_{\lambda}(F^1F^2;\vartheta,\varphi,\chi)]}^*_{\bar{\alpha}\bar{\beta}0}
      \nonumber \\
    {[h_{\lambda}(F^1F^2;\vartheta,-\varphi,-\chi)}]_{\alpha\bar{\beta} 0}
       & = &
    {-[h_{\lambda}(F^1F^2;\vartheta,\varphi,\chi)]}^*_{\bar{\alpha}\beta 0} .
\end{eqnarray}
\par
If the same vacuum is used on both sides, the above two types of
matrix elements are connected via
\begin{eqnarray}
  \label{symroth0abab0}
    {[h_{\lambda}(F^1F^1;\vartheta,-\varphi,-\chi)]}_{\alpha\beta 0}
       & = &
    (-1)^{K_{\alpha\beta}}
    {[h_{\lambda}(F^1F^1;\vartheta,\varphi,\chi)]}^*_{0\alpha\beta}
      \nonumber \\
    {[h_{\lambda}(F^1F^1;\vartheta,-\varphi,-\chi)]}_{\bar{\alpha}\bar{\beta}0}
       & = &
    (-1)^{K_{\alpha\beta}}
    {[h_{\lambda}(F^1F^1;\vartheta,\varphi,\chi)]}^*_{0\bar{\alpha}\bar{\beta}}
      \nonumber \\
    {[h_{\lambda}(F^1F^1;\vartheta,-\varphi,-\chi)]}_{\alpha\bar{\beta}0}
       & = &
    (-1)^{K_{\alpha\bar{\beta}}}
    {[h_{\lambda}(F^1F^1;\vartheta,\varphi,\chi)]}^*_{0\alpha\bar{\beta}}
      \nonumber \\
    {[h_{\lambda}(F^1F^1;\vartheta,-\varphi,-\chi)]}_{\bar{\alpha}\beta 0}
       & = &
    (-1)^{K_{\alpha\bar{\beta}}}
    {[h_{\lambda}(F^1F^1;\vartheta,\varphi,\chi)]}^*_{0\bar{\alpha}\beta} ,
\end{eqnarray}
which yields for 
$[h^{02}_{\lambda}(F^1F^1;\tilde{\Omega})]_{\alpha\beta}$ and
$[h^{20}_{\lambda}(F^1F^1;\tilde{\Omega})]_{\alpha\beta}$
\begin{eqnarray}
  \label{symroth0220}
    {[h^{02}_{\lambda}(F^1F^1;\vartheta,\varphi,\chi)]}_{\alpha\beta}
       & = &
    (-1)^{K_{\alpha\beta}}
  {[h^{20}_{\lambda}(F^1F^1;\vartheta,\varphi,\chi)]}_{\bar{\alpha}\bar{\beta}}
      \nonumber \\
    {[h^{02}_{\lambda}(F^1F^1;\vartheta,\varphi,\chi)]}_{\alpha\bar{\beta}}
       & = &
    - (-1)^{K_{\alpha\bar{\beta}}}
    {[h^{20}_{\lambda}(F^1F^1;\vartheta,\varphi,\chi)]}_{\bar{\alpha}\beta} .
\end{eqnarray}
The same relations hold for
$[\tilde{g}^{11}_{\lambda}(\tilde{\Omega})]_{\alpha\beta}$ and
$[g^{11}_{\lambda}(\tilde{\Omega})]_{\alpha\beta}$.
\par
For the rotated matrix elements 
\begin{eqnarray}
  \label{sym1rothabgd}
       \lefteqn{ [h_{\lambda}(F^1F^2;\vartheta,\varphi,\chi)]
                                           _{\alpha\beta\gamma\delta}
       \equiv } \nonumber \\
       & & \quad 
                    \langle F^1| a_{\beta}(F^1) a_{\alpha}(F^1)
                    \hat{H} \hat{\tilde{R}}(\vartheta,\varphi,\chi)
                    a^{\dagger}_{\gamma}(F^2) a^{\dagger}_{\delta}(F^2)
                    |F^2_{\lambda}\rangle
\end{eqnarray}
we get
\begin{eqnarray}
  \label{sym2rothabgd}
        {[h_{\lambda}(F^1F^2;\vartheta,-\varphi,-\chi)]}
                _{\alpha\beta\gamma\delta}
        & = &
        {[h_{\lambda}(F^1F^2;\vartheta,\varphi,\chi)]}^*
                _{\bar{\alpha}\bar{\beta}\bar{\gamma}\bar{\delta}}
        \nonumber \\
        {[h_{\lambda}(F^1F^2;\vartheta,-\varphi,-\chi)]}
                _{\alpha\beta\bar{\gamma}\bar{\delta}}
        & = &
        {[h_{\lambda}(F^1F^2;\vartheta,\varphi,\chi)]}^*
                _{\bar{\alpha}\bar{\beta}\gamma\delta}
        \nonumber \\
        {[h_{\lambda}(F^1F^2;\vartheta,-\varphi,-\chi)]}
                _{\alpha\beta\gamma\bar{\delta}}
        & = &
        -{[h_{\lambda}(F^1F^2;\vartheta,\varphi,\chi)]}^*
                _{\bar{\alpha}\bar{\beta}\bar{\gamma}\delta}
        \nonumber \\
        {[h_{\lambda}(F^1F^2;\vartheta,-\varphi,-\chi)]}
                _{\alpha\beta\bar{\gamma}\delta}
        & = &
        -{[h_{\lambda}(F^1F^2;\vartheta,\varphi,\chi)]}^*
                _{\bar{\alpha}\bar{\beta}\gamma\bar{\delta}}
        \nonumber \\
        {[h_{\lambda}(F^1F^2;\vartheta,-\varphi,-\chi)]}
                _{\alpha\bar{\beta}\gamma\delta}
        & = &
        -{[h_{\lambda}(F^1F^2;\vartheta,\varphi,\chi)]}^*
                _{\bar{\alpha}\beta\bar{\gamma}\bar{\delta}}
        \nonumber \\
        {[h_{\lambda}(F^1F^2;\vartheta,-\varphi,-\chi)]}
                _{\alpha\bar{\beta}\bar{\gamma}\bar{\delta}}
        & = &
        -{[h_{\lambda}(F^1F^2;\vartheta,\varphi,\chi)]}^*
                _{\bar{\alpha}\beta\gamma\delta}
        \nonumber \\
        {[h_{\lambda}(F^1F^2;\vartheta,-\varphi,-\chi)]}
                _{\alpha\bar{\beta}\gamma\bar{\delta}}
        & = &
        {[h_{\lambda}(F^1F^2;\vartheta,\varphi,\chi)]}^*
                _{\bar{\alpha}\beta\bar{\gamma}\delta}
        \nonumber \\
        {[h_{\lambda}(F^1F^2;\vartheta,-\varphi,-\chi)]}
                _{\alpha\bar{\beta}\bar{\gamma}\delta}
        & = &
        {[h_{\lambda}(F^1F^2;\vartheta,\varphi,\chi)]}^*
                _{\bar{\alpha}\beta\gamma\bar{\delta}} .
\end{eqnarray}
For identical vacua on both sides furthermore
\begin{equation}
  \label{sym3rothabgd}
        {[h_{\lambda}(F^1F^1;\vartheta,-\varphi,-\chi)]}
                _{\alpha\beta\gamma\delta}
        =
        (-1)^{K_{\alpha\beta}+K_{\gamma\delta}}
        {[h_{\lambda}(F^1F^1;\vartheta,\varphi,\chi)]}^*
                _{\gamma\delta\alpha\beta} 
\end{equation}
holds.
\par
Finally, for the Hamiltonian matrix elements in between two
one-quasiparticle configurations
\begin{equation}
  \label{sym1rotab}
       [h_{\lambda}(F^1F^2;\vartheta,\varphi,\chi)]_{\alpha\beta}
       \equiv
       \langle F^1| a_{\alpha}(F^1)
                    \hat{H} \hat{\tilde{R}}(\vartheta,\varphi,\chi)
                    a^{\dagger}_{\beta}(F^2)
              |F^2_{\lambda}\rangle
\end{equation}
the relations
\begin{eqnarray}
  \label{sym2rothab}
        {[h_{\lambda}(F^1F^2;\vartheta,-\varphi,-\chi)]}_{\alpha\beta}
        & = &
        {[h_{\lambda}(F^1F^2;\vartheta,\varphi,\chi)]}^*
                                             _{\bar{\alpha}\bar{\beta}}
        \nonumber \\
        {[h_{\lambda}(F^1F^2;\vartheta,-\varphi,-\chi)]}_{\alpha\bar{\beta}}
        & = &
        - {[h_{\lambda}(F^1F^2;\vartheta,\varphi,\chi)]}^*
                                             _{\bar{\alpha}\beta} 
\end{eqnarray}
are found.
\end{appendix}
\newpage
\newpage
\section*{Figure Captions}
\begin{description}
\item[Fig.~1]
The energy spectrum for the nucleus $^{20}$Ne as obtained by three 
different approaches~: the complete shell model diagonalization 
(SCM, solid lines), the {\sl complex} MONSTER (CM, dashed lines) and 
the {\sl real} MONSTER (RM, dotted lines). The energy is given relative to 
the $^{16}$O core. Only excitation energies up to 16 MeV above the shell 
model ground state are presented. The {\sl real} MONSTER calculation
has been based on the corresponding $0^+$ VAMPIR transformation.
The {\sl complex} MONSTER results, which do reproduce all shell model
states exactly, are independent of the particular choice of the underlying
transformation.
\item[Fig.~2]
Same as Fig.~1, but for $^{22}$Ne. For each spin only the lowest shell model
states are presented. The {\sl complex} MONSTER was here built on the
{\sl complex} VAMPIR solution obtained for the $0^+$ ground state.
\item[Fig.~3]
The energies of the yrast states in $^{22}$Ne as obtained by various 
approaches~: the shell model (SCM), the {\sl complex} MONSTER (CM)
(again on top of the $0^+$ VAMPIR solution), the {\sl complex} VAMPIR (CV),
the {\sl real} MONSTER (RM) (based on the corresponding $0^+$ VAMPIR solution, 
too), and the {\sl real} VAMPIR (RV). In the last approach only the
even spin states are accessible. Spin and parity are indicated 
on the l.h.s.~of each level. Again the energy is given relative
to the $^{16}$O core.
\item[Fig.~4]
The five lowest $0^+$ MONSTER states of $^{22}$Ne are plotted versus
the spin of the VAMPIR transformation which was used in each calculation.
Solid lines refer to the {\sl complex} MONSTER, dotted lines to the 
{\sl real} MONSTER results. In the latter case only transformations
with even spin values could be used. For comparison the lowest three 
$0^+$ shell model (SCM) energies are also given. Again the energy is 
given relative to the $^{16}$O core.
\end{description}
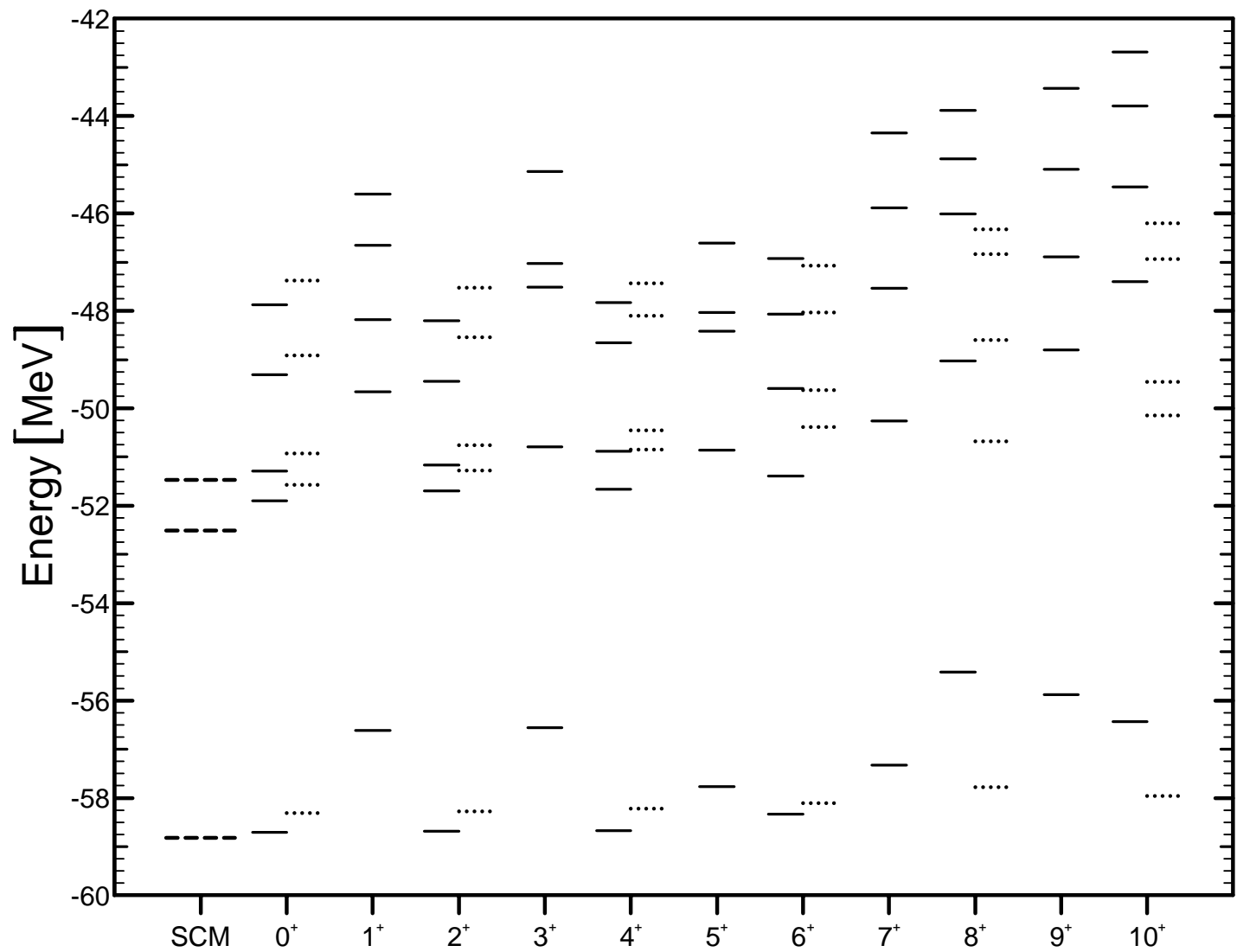
\end{document}